\documentclass[fleqn,usenatbib]{mnras}

\usepackage{newtxtext,newtxmath}
\usepackage{orcidlink} 
\usepackage{bm}
\usepackage{color}
\usepackage{caption}
\usepackage{float}
\usepackage{subcaption}
\usepackage{stfloats}
\usepackage{pifont}
\usepackage{enumitem} 

\newcommand\scalemath[2]{\scalebox{#1}{\mbox{\ensuremath{\displaystyle #2}}}}

\makeatletter
\def\fps@figure{!htbp}
\def\fps@table{htbp}
\makeatother

\usepackage[T1]{fontenc}
\usepackage{xcolor}

\DeclareRobustCommand{\VAN}[3]{#2}
\let\VANthebibliography\thebibliography
\def\thebibliography{\DeclareRobustCommand{\VAN}[3]{##3}\VANthebibliography}

\usepackage{graphicx}	
\usepackage{amsmath}	
\usepackage{float}

\usepackage{hyperref}
\hypersetup{
    colorlinks=true,
    linkcolor=black,  
    citecolor=blue,  
    filecolor=black,  
    urlcolor=blue    
}

\usepackage[acronym,toc]{glossaries-extra}
\usepackage[normalem]{ulem}
\title[The $k$-e$\mu$lator]{$k$-e$\mu$lator: emulating clustering effects of the $k$-essence dark energy }

\author[A. R.~Nouri-Zonoz, F.~Hassani, M.~Kunz]{
A. R. Nouri-Zonoz,$^{1}$ \orcidlink{0009-0006-6164-8670}\thanks{E-mail: Ahmadreza.Nourizonoz@unige.ch}
Farbod Hassani,$^{2}$ \orcidlink{0000-0003-2640-4460}
Martin Kunz$^{1}$ \orcidlink{0000-0002-3052-7394}
\\
$^{1}$Universit\'e de Gen\`eve, D\'epartement de Physique Th\'eorique and CAP, Gen\`eve 4 , Switzerland\\
$^{2}$Institute of Theoretical Astrophysics, University of Oslo, 
0315 Oslo, Norway\\
}

\date{Accepted XXX. Received YYY; in original form ZZZ}

\pubyear{2024}

\begin{document}
\graphicspath{ {./Figures/} }
\label{firstpage}
\pagerange{\pageref{firstpage}--\pageref{lastpage}}
\maketitle

\begin{abstract}
We build an emulator based on the polynomial chaos expansion (PCE) technique to efficiently model the non-linear effects associated with the clustering of the $k$-essence dark energy in the effective field theory (EFT) framework. These effects can be described through a modification of Poisson's equation, denoted by the function $\mu(k,z)$, which in general depends on wavenumber $k$ and redshift $z$. To emulate this function, we perform $200$ high-resolution $N$-body simulations sampled from a seven-dimensional parameter space with the Latin hypercube method. These simulations are executed using the \texttt{k-evolution} code on a fixed mesh, containing $1200^3$ dark matter particles within a box size of $400~\text{Mpc}/ h$. The emulation process has been carried out within \texttt{UQLab}, a \texttt{MATLAB}-based software specifically dedicated to emulation and uncertainty quantification tasks. Apart from its role in emulation, the PCE method also facilitates the measurement of Sobol indices, enabling us to assess the relative impact of each cosmological parameter on the $\mu$ function.
Our results show that the PCE-based emulator efficiently and accurately reflects the behavior of the $k$-essence dark energy for the cosmological parameter space defined by $w_0 c_s^2 \text{CDM} +\sum m_{\nu}$. Compared against actual simulations, the emulator achieves sub-percent accuracy up to the wavenumber $k \approx 9.4 ~h \text{Mpc}^{-1} $ for redshifts $z \leq 3$.  Our emulator provides an efficient and reliable tool for Markov chain Monte Carlo (MCMC) analysis, and its capability to closely mimic the properties of the $k$-essence dark energy makes it a crucial component in Bayesian parameter estimations. The code is publicly available at \url{https://github.com/anourizo/k-emulator}.


\end{abstract}




\section{Introduction}
Based on the theory of general relativity, the $\Lambda$CDM cosmological model, which incorporates the collisionless cold dark matter (CDM) and the cosmological constant $\Lambda$ associated with dark energy, has long been the standard paradigm in describing the large-scale structure, evolution, and the late-time accelerating expansion of the Universe. Strengthened by its remarkable consistency with diverse observational data, from the analysis of cosmic microwave background (CMB) power spectrum \citep{Planck:2015bpv} to the large-scale galaxy clustering surveys \citep{BOSS:2016wmc}, it has earned widespread acceptance within the cosmology community.  However, despite the many triumphs of the model, the characterization of the cosmological constant as dark energy has encountered multiple challenges. Among them,
the cosmological constant problem \citep{Adler:1995vd} and the cosmic coincidence problem \citep{CRPHYS_2012__13_6-7_566_0} stand as notable issues within the $\Lambda$CDM cosmological framework. Faced with these persistent issues and several tensions within the $\Lambda$CDM context,  particularly the divergent measurements of the Hubble constant ($H_0$) and the structure growth parameter ($S_8$) between \textit{Planck} satellite data \citep{Planck:2018vyg} and local observations—such as supernova analyses for $H_0$ \citep{Riess:2020fzl} and weak lensing experiments for $S_8$ \citep{Hildebrandt:2016iqg, Hildebrandt:2018yau, DES:2017myr, KiDS:2020suj}— the investigation and test of modified gravity and dark energy theories has become increasingly relevant. These pursuits are now more feasible than ever, thanks to the advancements in cosmological instrumentation. Such developments have brought us into a crucial phase of research, allowing us to constrain different dark energy candidates with an unprecedented accuracy.
At the forefront of these instruments, future and ongoing surveys such as Euclid\footnote{\url{https://sci.esa.int/web/euclid}} \citep{EUCLID:2011zbd}, the Vera C. Rubin observatory\footnote{\url{https://rubinobservatory.org/}} \citep{LSSTScience:2009jmu}, DESI\footnote{\url{https://www.desi.lbl.gov/}} \citep{DESI:2016fyo}, and the Nancy Grace Roman Space Telescope\footnote{\url{https://science.nasa.gov/mission/roman-space-telescope/}} \citep{Akeson:2019biv} will allow us to shrink the error bars of DE parameters  through the combined use of galaxy clustering and weak lensing methods.
This will be of particular interest in the $k$-essence dark energy models \citep{Armendariz-Picon:2000nqq} where the current data offers weak constraints on the speed of sound \citep[see e.g.][]{Planck:2015bue}.

As the Universe evolves, the growth of structures on small scales leads to non-linear structure formation in cosmic evolution where linear perturbation theory is no longer applicable. 
This complexity has made it challenging to simply rely on traditional theoretical approaches, especially considering that even the effective field theory of large scale structures (EFTofLSS) fails beyond $k \sim 0.6 ~\text{h/Mpc}$ \citep{Carrasco:2013mua}. Within this landscape, $N$-body simulations have emerged as indispensable tools. In particular, the \texttt{gevolution} code \citep{Adamek:2016zes} is the first $N$-body simulation for cosmic structure formation that is based on
general relativity employing the weak field approximation. This code is structured in such a way that facilitates the implementation of dark energy and modified gravity theories. An example is the implementation of the $k$-essence dark energy model in the form of the \texttt{k-evolution} code \citep{Hassani:2019lmy}.

Yet, while $N$-body simulations offer a reliable tool to test different models of gravity/dark energy, one cannot ignore the significant computational demands they impose. The high resolution requirements for accurate predictions makes these simulations extremely resource-intensive, both in terms of computational power and time. Running multiple simulations for parameter exploration can quickly become infeasible, especially in the context of Bayesian best fit parameter analysis, where thousands of model evaluations are typically necessary. This computational challenge has given rise to an increasing interest in the development and application of emulators. 

Emulators are fast and accurate approximations of full simulations, trained on a subset of the parameter space. As surrogate models, they can quickly provide estimates for different parameter sets without bearing the substantial computational expense, making them ideal for extensive parameter searches. Their ability to boost the efficiency of likelihood sampling is crucial for the optimal use of Markov chain Monte Carlo (MCMC) techniques. Examples of cosmological emulators include \texttt{\texttt{EuclidEmulator}} \citep{Euclid:2018mlb, Euclid:2020rfv}, \textit{Aemulus} project \citep{DeRose:2018xdj, McClintock:2018uyf, Zhai:2018plk, Storey-Fisher:2022etc}, \texttt{FrankenEmu} \citep{Heitmann:2013bra}, \texttt{CosmicEmu} \citep{Heitmann:2009cu, Heitmann:2008eq, Lawrence:2009uk, Heitmann:2015xma, Lawrence:2017ost, Moran:2022iwe}, \texttt{BE-HaPPY} \citep{Valcin:2019fxe}, \texttt{NGenHalofit} \citep{Smith:2018zcj}, \texttt{Dark quest} emulator \citep{Nishimichi:2018etk}, \texttt{CosmoPower} \citep{SpurioMancini:2021ppk} and \texttt{CosmoPower-JAX} \citep{Piras23}.

 In this paper we introduce a new emulator for the $k$-essence dark energy model based on \texttt{k-evolution} simulations, assuming constant values for the equation of state parameter $w_0$ and the speed of sound $c_s$. The key aim of this emulator is to accurately reproduce the clustering effects of the $k$-essence dark energy for small values of sound speed, which is a distinct feature of the \texttt{k-evolution} code. As we show later, when the speed of sound is large, the results obtained from Boltzmann codes remain reliable.

In the construction of the emulator, we use sparse polynomial chaos expansion (SPCE) method, as described in detail in \cite{BlatmanThesis}. This process is conducted using \texttt{UQLab}, a specialized software for uncertainty quantification (UQ) introduced in \cite{Marelli2014UQLab}.
The effectiveness of  SPCE is not only highlighted in our work but has also been prominently featured in the publications of the \texttt{\texttt{EuclidEmulator}} \citep{Euclid:2018mlb, Euclid:2020rfv}.

This paper is organized as follows: In Section \ref{sec:theoretical framework} we discuss the theoretical framework of the $k$-essence dark energy and explain its clustering impact on Poisson's equation through the $\mu$ function, which is the target of the our emulator. 
In Section \ref{sec:EmuStages}, we outline the main steps of the emulator construction and discuss each one in detail.
Following this, in Section \ref{sec:fine-tuning}, we explain the fine-tuning procedure of parameters involved in the construction of the emulator, aiming to minimize emulation errors. Section \ref{sec:Sobol} is dedicated to the sensitivity analysis conducted using Sobol indices with the purpose of identifying the key parameters influencing the emulator's target, in this case the $\mu$ function. In Section \ref{sec:simulation-based emulator}, we transfer the knowledge obtained from the mock emulator (\texttt{Halofit}-based emulator) to the development of a simulation-based emulator. This section also covers discussions on  convergence tests of simulations and preprocessing of the training data. Finally, in Section \ref{sec:reproduce}, we explain the application of our emulator in obtaining the gravitational potential power spectrum using the matter power spectrum. The paper concludes with a summary of our results.


\section{Theoretical framework}
\label{sec:theoretical framework}

\subsection{The {\boldmath{$k$}}-essence dark energy model}
$k$-essence models are a family of theories based on a scalar field where a generic non-canonical Lagrangian term, often denoted as $P(\phi,X)$, governs the scalar field's dynamics. Here $\phi$ is the $k$-essence scalar field  and $X = -\frac{1}{2}g^{\mu \nu}\partial_{\mu}\phi\partial_{\nu}\phi$ is the kinetic component. The action for the $k$-essence models is written as 
\begin{equation}
    S = \int d^4x \sqrt{-g} \left( P(\phi, X) + \frac{R}{16\pi G} +\mathcal{L}_m \right),
	\label{eq:action}
\end{equation}
where $g$ is the determinant of the metric $g_{\mu\nu}$, $P(\phi,X)$ is the generic form of the $k$-essence Lagrangian, $R$ is the Ricci scalar and $\mathcal{L}_m $ is the matter Lagrangian. This form of action was first proposed by \citet{Armendariz-Picon:1999hyi} as an alternative to the standard slow-roll scenario, describing the  inflationary phase of the Universe. Later on, the scheme's initial adaptation to dark energy was studied in \citet{Chiba:1999ka} and \citet{Armendariz-Picon:2000nqq}.
\\
To study inhomogeneities in the Universe it is common to adopt the Friedman-Lema\^itre-Robertson-Walker (FLRW) metric in the Poisson gauge which reads
\begin{equation}
    ds^2 = a^2(\eta)[-e^{2\Psi}d\eta^2-2B_idx^id\eta + (e^{-2\Phi}\delta_{ij}+h_{ij})dx^idx^j],
	\label{eq:FLRW}
\end{equation}
where $a(\eta)$ is the scale factor, $d\eta = dt/a(t)$ is the differential element of conformal time, $\Psi$ and $\Phi$ denote the temporal and spatial scalar perturbations, and $B_i$ and $h_{ij}$ represent the transverse vector ($\partial^i B_i = 0$) and transverse-traceless tensor ($\partial^i h^j_i = h^i_i = 0$) perturbations, were we have used the Einstein summation convention over the repeated indices.

From the action defined in (\ref{eq:action}), the stress-energy tensor of the $k$-essence scalar field is calculated as,
\begin{equation}
\begin{aligned}
     T^{(\phi)}_{\mu\nu} &= -\frac{2}{\sqrt{-g}}\frac{\delta (\sqrt{-g} P)}{\delta g^{\mu\nu}} \\
     & = -\frac{2}{\sqrt{-g}} \left( \sqrt{-g}\frac{\partial P}{\partial X}\frac{\delta X}{\delta g^{\mu\nu}} + \frac{\delta \sqrt{-g}}{\delta g^{\mu\nu}} P \right)\\
     & = -\frac{2}{\sqrt{-g}} \left(-\frac{\sqrt{-g}}{2} P _{, X} \partial_\mu \phi \partial_\nu \phi - \frac{\sqrt{-g}}{2} g_{\mu\nu} P \right)  \\
     & = P_{,X}\partial_\mu{\phi}\partial_\nu{\phi}+ g_{\mu\nu}P .
	\label{eq:stress-energy-k-essence}
 \end{aligned}
\end{equation}
By setting
\begin{equation}
\rho_\phi = 2XP_{,X} -P, \qquad P_\phi = P, \qquad u_\mu = \frac{\partial_\mu \phi}{\sqrt{2X}},
    \label{eq:effective-quantities}
\end{equation}
it is possible to express the given stress-energy tensor above in the form of a perfect fluid, $ T_{\mu\nu} = (\rho_{\phi} +P_{\phi})u_\mu u_\nu + g_{\mu \nu} P_{\phi}$.
Following the expressions in (\ref{eq:effective-quantities}), the stress-energy tensor of the $k$-essence fluid, which lacks anisotropic stress ($\sigma$), can be characterised by its equation of state $w_\phi$ and squared speed of sound $c_s^2$ \citep[for more details see][]{Hassani:2019lmy}
\begin{equation}
w_{\phi} = \frac{P_\phi}{\rho_\phi} = \frac{P}{2XP_{,X} -P} \quad,\quad c_s^2 = \left(\frac{\partial P}{\partial \rho}\right)\bigg|_{\phi}= \frac{P_{,X}}{P_{,X}+ 2XP_{,XX}},
    \label{eq:w-cs2}
\end{equation}
where the comma stands for the partial derivative {while the symbol $|_{\phi}$ denotes the derivative when $\phi$ is held constant. Note that within the context of $k$-essence models, both the equation of state parameter, $w_\phi$, and the speed of sound, $c_s$, are time-dependent in general. However, for the purposes of our analysis, we will treat them as constants. 
Allowing the equation of state parameter and the speed of sound to vary with time leads to an increase in the number of free parameters, which complicates both the theoretical framework and the numerical analysis. Moreover, from the theoretical point of view, one can parametrize the time evolution of $w_{\phi}, c_s^2$ in different ways by considering assumptions for the function $P(X, \phi)$.
This topic is interesting in itself and should be considered in later investigations. To sidestep these complexities, which require detailed analysis, we have chosen to defer the exploration of time-varying parameters to future studies.
Generally, when it comes to studying the behaviour of the $k$-essence dark energy, both $c_s^2$ and $w_{\phi}$ are essential parameters to look into (hereafter, we will use $w_0$ instead of $w_{\phi,0}$ to represent today’s value of the equation of state parameter).
It is also worth noting that in our study we use the total stress-energy tensor which includes contributions from the matter sector and the $k$-essence field
\begin{equation}
T_{\mu\nu} = T_{\mu\nu}^{(m)} + T_{\mu\nu}^{(\phi)}.
\label{full-stress}
\end{equation}

\subsection{Impact of dark energy perturbations on Poisson's equation}

We can obtain the gravitational field equations by variying the action (\ref{eq:action}) with respect to the metric. In the weak field regime,  one equation is the Hamiltonian constraint, which can be seen as a relativistic Poisson equation  \citep{Hassani:2019wed},
\begin{equation}
    \nabla^2 \Phi = 4\pi G a^2 \sum_X\bar\rho_X\Delta_X + S \, ,
	\label{eq:Poisson}
\end{equation}
where
\begin{equation}
    S \equiv \frac{1}{2}\delta^{ij}\Phi_{,i} \Phi_{,j} - 8\pi G a^2\Phi \sum_X\bar\rho_X\Delta_X
	\label{eq:Short-wave}
\end{equation}
is called the shortwave correction and is due to non-linear contributions in the weak field expansion. As shown in \citet{Hassani:2019wed}, the shortwave correction is small and can safely be ignored. $\bar \rho_X$ and $\Delta_X$ are the background density and the comoving density contrast for each species including cold dark matter, baryon and $k$-essence dark energy (as our focus is on the late time stage of the Universe, we will neglect the contribution from radiation). 
Switching to Fourier space, contributions to the $\Phi$ metric perturbation from a clustering dark energy component like $k$-essence
can be captured through a function $\mu(k,z)$,
 \begin{equation}
    -k^2 \Phi =  4\pi G a^2 \mu(k,z) \sum_{X\backslash _\text{DE}}\bar\rho_X\Delta_X \, ,
	\label{eq:modified-poisson}
\end{equation}
where the symbol $X\backslash _\text{DE}$ means the summation excludes the contribution from the dark energy component. This formalism is more general and can capture any deviations from $\Lambda$CDM, it is for example also applicable to modified-gravity models  \citep{Amendola:2016saw}. Generally speaking, the $\mu$ function tells us how much the Poisson equation is modified as a result of dark energy perturbations or modification of General Relativity. Within the $\Lambda$CDM framework where dark energy perturbations are absent, one would obtain the standard Poisson equation, which implies $\mu(k,z) = 1$ for all scales and redshifts. 

As already mentioned, at late times and on small scales the sum in Eq.\ \eqref{eq:modified-poisson} is dominated by the contribution from matter so that we can simply set $X=m$. This suggests defining the $\mu$ function as
\begin{equation}
\mu(k,z) = 1 + \frac{\bar\rho_{\text{DE}}(z)\Delta_{\text{DE}}(k,z)}{\bar\rho_m(z)\Delta_m(k,z)} \, .
    \label{mu-1}
\end{equation}
However, in actual observations we
will in general be considering power spectra of the metric and matter density perturbations, and for this purpose we define a function $\mu^2$ through the expression
\begin{equation}
\mu^2(k,z) = \frac{k^4 {\langle}\Phi\Phi^* {\rangle}}{(4\pi G a^2\bar \rho_m)^2\langle\Delta_m \Delta_m^*\rangle} \, .
    \label{mu-0}
\end{equation}
Naively we might think that we can simply replace $\Delta_{\text{DE}}(k,z)$ and $\Delta_m(k,z)$ in Eq.\ \eqref{mu-1} with the square root of the dark energy and matter power spectra. However, inserting the relativistic Poisson equation (keeping only matter and dark energy)
\begin{equation}
    -k^2 \Phi = 4\pi G a^2 \left( \bar\rho_{\rm DE} \Delta_{\rm DE} +\bar\rho_m\Delta_m \right)
	\label{eq:Poisson_de_m}
\end{equation}
into Eq.\ \eqref{mu-0}, we obtain
\begin{equation}
\begin{aligned}
   \mu^2(k,z)&=  \langle \mu(k,z) \mu(k,z)^* \rangle \\ 
    & =\frac{\bar{\rho}^2_\text{DE}\langle\Delta^{}_\text{DE}\Delta_\text{DE}^*\rangle+
\bar{\rho}^2_m\langle\Delta^{}_m\Delta^*_m\rangle+ 2\bar{\rho}_\text{DE}\bar{\rho}_m\langle\Delta^{}_\text{DE}\Delta^{*}_m\rangle}{ \bar{\rho}^2_m\langle\Delta^{}_m\Delta^{*}_m\rangle} \, .
\label{eq:mu-corr}
\end{aligned}
\end{equation}
If the dark energy and dark matter are fully correlated such that the normalized cross-correlation coefficient $f_\times$, defined as
\begin{equation}
f_\times = \frac{\langle\Delta^{}_\text{DE}\Delta^{*}_m\rangle}{\sqrt{\langle\Delta^{}_\text{DE}\Delta_\text{DE}^*\rangle\langle\Delta^{}_m\Delta^*_m\rangle}}
\end{equation}
is equal to unity on all scales, then the $\mu$ defined in Eq.\ \eqref{mu-1} can be recovered by taking the square root of the  Eq.\ \eqref{eq:mu-corr}. In linear perturbation theory this is indeed the case, but on non-linear scales the situation is less straightforward and the value of the speed of sound can significantly influence this correlation. We will discuss this in more details in the following subsection and in Section \ref{sec:reproduce}.


\subsection{Numerical result from the {\boldmath{$k$}}-evolution code}

\begin{figure*}
    \centering
    \includegraphics[width=0.9\textwidth]{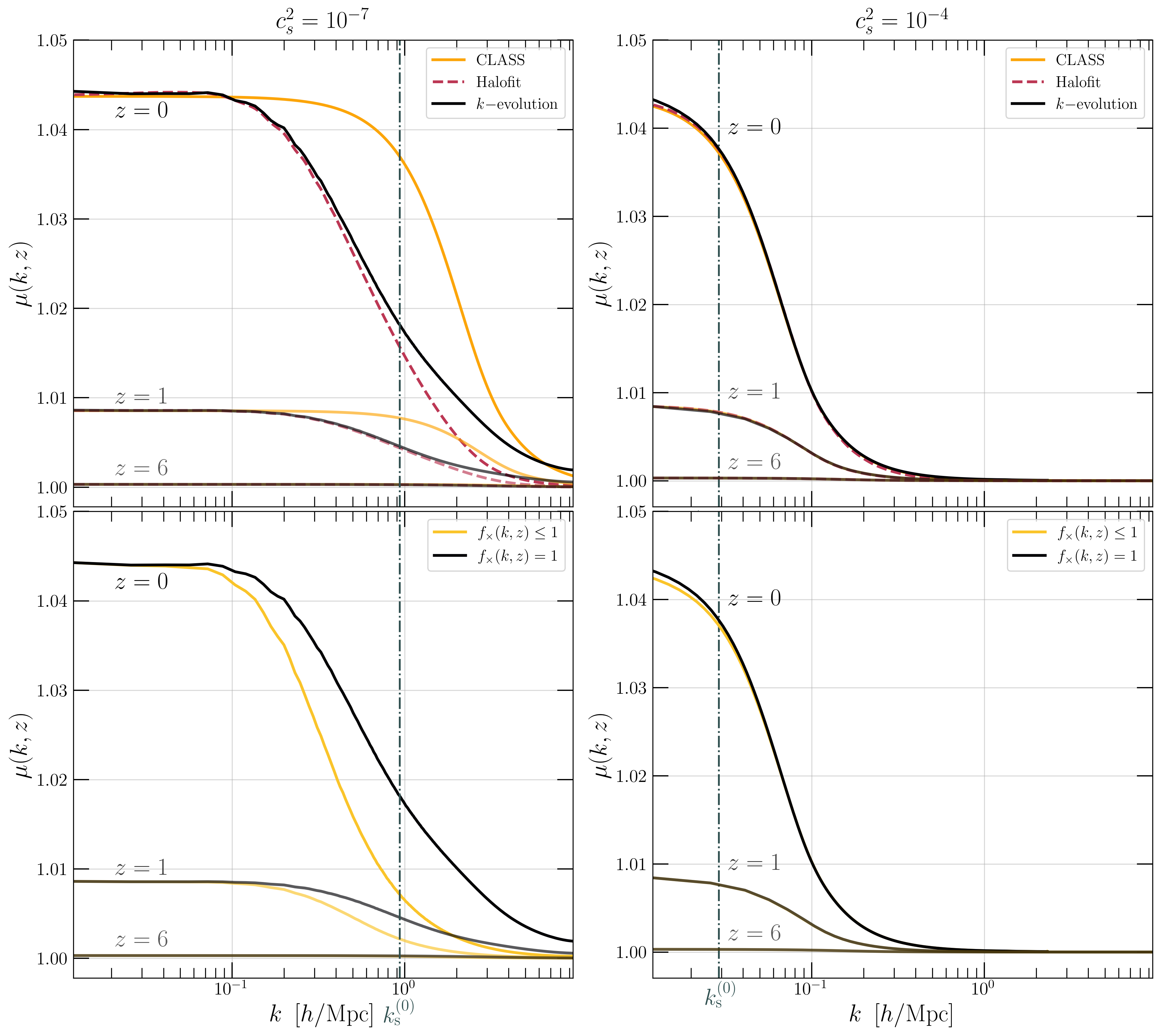}
    \caption{ \textbf{Upper panels} : Comparison of the $\mu$ function for two cases of $c_s^2 = 10^{-7}$ (left panel) and $c_s^2 = 10^{-4}$ (right panel) and $w_0 = -0.9$, obtained from \texttt{CLASS}, \texttt{Halofit} and \texttt{k-evolution} at $z = 0,1,6$. In both cases, by moving to higher redshifts and smaller scales, the effect of $k$-essence perturbations decreases; the former is due to the sub-dominant contribution of dark energy in the past compared to matter component and the latter is due to the existence of sound horizons. For $c_s^2 = 10^{-4}$ at $z = 0$, the sound horizon wavenumber is positioned at $k_s \approx 0.029 ~\text{h/Mpc}$, beyond which, dark energy perturbations start to decay. As this scale is larger than the matter non-linearity threshold, linear perturbation theory remains sufficient for the prediction of density contrasts. Consequently, the results from \texttt{CLASS}, \texttt{Halofit}, and \texttt{k-evolution} align when transitioning from large to small scales. For the case when $c_s^2 = 10^{-7}$ at $z=0$ , the sound horizon is located at $k_s \approx 0.94 ~\text{h/Mpc}$ which coincides with the matter non-linearity scale. After entering this region the result from \texttt{CLASS} separates from those of \texttt{Halofit} and \texttt{k-evolution} as the latter two account for the non-linear evolution of matter (while the dark energy perturbation grows slower compared to matter), leading them to predict lower values for $\mu$. At smaller scales, when dark energy undergoes non-linear evolution, a deviation is observed between the results of \texttt{k-evolution} and \texttt{Halofit}. This difference is attributed to the non-linear evolution of $\Delta_\text{DE}$ in \texttt{k-evolution}, which in turn yields higher values for the $\mu$ function compared to the \texttt{Halofit}. \\
    \textbf{Lower panels} : Comparison of the $\mu$ function in two scenarios: $c_s^2 = 10^{-7}$ (left panel) and $c_s^2 = 10^{-4}$ (right panel), with $w_0 = -0.9$. The results are derived from $k$-evolution, taking into account both the actual cross-correlation $f_\times(k,z) \leq 1$ and the assumption that $f_\times(k,z) = 1$. For $c_s^2 = 10^{-4}$, the agreement of $\mu$ functions across all scales in the two mentioned cases is attributed to dark energy and matter perturbations being well described by the linear theory and consequently being fully correlated with each other. In contrast, in the case of $c_s^2 = 10^{-7}$, the transition of the $\mu$ in which $f_\times(k,z)\leq 1$  is suppressed earlier compared to the scenario where $f_\times(k,z) = 1$, underscoring the significant influence of cross-correlation on the behavior of $\mu$ for small speeds of sound. For the \texttt{k-evolution} data, we ran two simulations, each evolving $3840^3$ particles within boxes of comoving lengths $1280~\text{Mpc}/h$ and $9000~\text{Mpc}/h$.}
    \label{fig:mu_comparison}
\end{figure*}

\begin{figure*}
    \centering
    \includegraphics[width=0.95\textwidth]{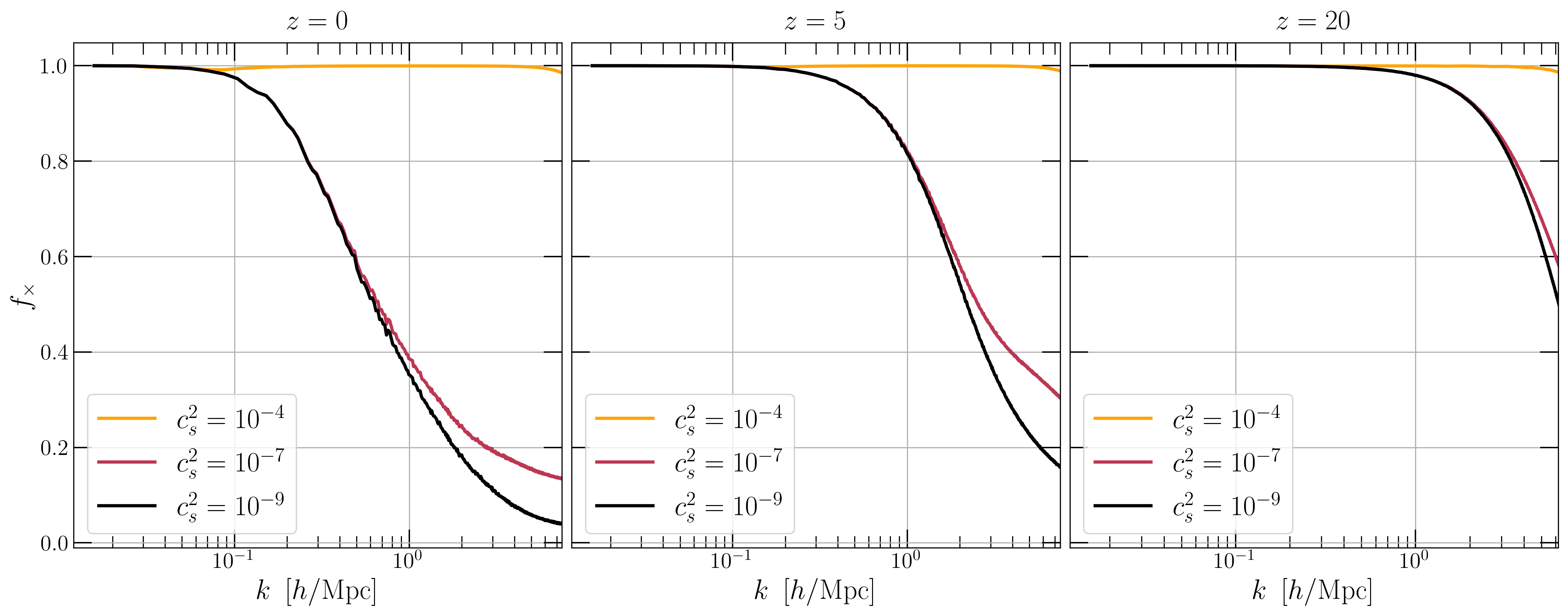}
    \caption{Normalized cross-correlation power spectra at $z = 0,5,20$, for the cases of  $c_s^2 = 10^{-4}$, $c_s^2 = 10^{-7}$ and $c_s^2 = 10^{-9}$. For the lower speeds of sound, $c_s^2 = 10^{-7}$ and $c_s^2 = 10^{-9}$, where dark energy clusters and has its own dynamics, there is a noticeable decline in cross-correlation coefficient on small scales. In contrast, at the larger sound speed of $c_s^2 = 10^{-4}$, dark energy and matter densities maintain a full correlation across almost all scales.}
    \label{fig:corr_fact}
\end{figure*}

The \texttt{k-evolution}\footnote{\url{https://github.com/FarbodHassani/k-evolution}} code is an extension of the relativistic $N$-body code \texttt{gevolution}\footnote{\url{https://github.com/gevolution-code/gevolution-1.2}} \citep{Adamek:2016zes} in which the  effective field theory (EFT) approach to the $k$-essence dark energy has been implemented, see \citet{Hassani:2019lmy} for an in-depth discussion}. A distinguishing feature of \texttt{k-evolution} compared to \texttt{gevolution} is that it fully couples the $k$-essence equation of motion to the non-linear evolution of matter. In contrast, \texttt{gevolution} employs the linear solution provided by the Boltzmann code \texttt{CLASS} \citep{Blas:2011rf} to compute the dark energy perturbations. 

We illustrate the importance of including non-linearities in Fig.\ \ref{fig:mu_comparison} where we show $\mu(k,z)$ function obtained from Eq.\ \eqref{eq:mu-corr} for both cases where we naively assume $f_\times = 1$ and more realistic case of $f_\times \leq 1$. We compare $\mu(k,z)$ function at different redshifts using \texttt{CLASS} (linear perturbation theory only), \texttt{Halofit} (the matter power spectrum is non-linear) \citep{Takahashi:2012em} and \texttt{k-evolution} codes.
The cosmological parameters used in these comparisons are listed in Table \ref{table:cosmoparams} and include

\begin{itemize}
    \item  $w_0$: the dark energy equation-of-state (EoS)
    \item  $h$: the dimensionless Hubble parameter 
    \item $\Omega_\text{b}$: fractional energy density of baryonic matter 
    \item $\Omega_{\text{cdm}}$: fractional energy density of cold dark matter 
    \item $A_s$: spectral amplitude 
    \item $n_s$: spectral index 
\end{itemize}
We examine these parameters for two specific values of the speed of sound squared, $c_s^2 = 10^{-7}$ and $c_s^2 = 10^{-4}$.

\begin{table}
	\centering
	\caption{Cosmological parameters used in the computation of $\mu$ in Fig. \ref{fig:mu_comparison}.}
	\label{table:cosmoparams}
	\begin{tabular}{|c|c|c|c|c|c|} 
		\hline
		$w_0$ & h & $\Omega_\text{b}$ & $\Omega_{\text{cdm}}$ & $A_s$ & $n_s$ \\
		\hline\hline
		 -0.90 & 0.6755 & 0.0486 & 0.2617 & $2.21 \times 10^{-9}$ & 0.9660 \\
		\hline
	\end{tabular}
\end{table}


As discussed e.g.\ in \cite{Sapone:2009mb}, on large scales, outside the sound horizon, the dark energy clusters proportionally to the dark matter, with a relative factor that depends primarily on $w_0$. Within the sound horizon, the dark energy perturbations are suppressed. The overall shape that we expect for $\mu$ in a $k$-essence cosmology is a constant value on large scales, and then a transition at the sound horizon, with $\mu$ approaching $1$ for large values of $k$. This overall shape is clearly visible in Fig. \ref{fig:mu_comparison}. We see that the plateau on large scales has a value of about $1.043$ today for our case of $w_0=-0.9$. At higher redshift, this value scales with the relative evolution of the dark energy and dark matter densities, $(1+z)^{3w_0}$ in our model where $w_0$ is constant. 

 
 
 In the case of $c_s^2 = 10^{-4}$ at $z=0$, the sound horizon is located at $k_s \approx 0.029 ~\text{h/Mpc}$, where we can see a clear cut-off in $\mu$ function. 
 Given that this horizon is significantly larger than the matter non-linearity scale, $k_\text{nl}^{m} \approx 0.1 ~\text{h/Mpc}$ at $z=0$, the transition in $\mu$ is the same for \texttt{CLASS},  \texttt{Halofit} and \texttt{k-evolution} as both $\Delta_\text{DE}, \Delta_m$ can be perfectly described by linear perturbation theory. In the case of $c_s^2 = 10^{-7}$ however, the sound horizon is at $k_s = 0.94 ~\text{h/Mpc}$ at $z=0$, which falls within the scale where matter undergoes non-linear evolution. Upon entering the scale of matter non-linearity, $\Delta_m$ starts to increase faster than in the linear solution, while the dark energy density contrast, $\Delta_\text{DE}$, does not increase as significantly. This leads to a decrease in the $\mu$ function. Consequently, in this region, \texttt{Halofit} and \texttt{k-evolution} yield results that are lower than \texttt{CLASS} as $\Delta_m^\text{nl}> \Delta_m^\text{lin}$. Both \texttt{Halofit} and \texttt{k-evolution} continue to follow the same pattern until they approach the phase where dark energy begins its non-linear evolution.
In this phase, the predictions of \texttt{k-evolution} start to deviate from those of \texttt{Halofit}, indicating higher values for $\mu$. This stems from the \texttt{k-evolution}'s ability to include $k$-essence clustering as it gets drawn into the dark matter potential. Such clustering leads to an increase in $\Delta_\text{DE}$, and subsequently to an increase in the $\mu$ function $(\Delta_\text{DE}^\text{nl}> \Delta_\text{DE}^\text{lin})$. 

\begin{figure*}
    \centering
    \includegraphics[width=1\textwidth]{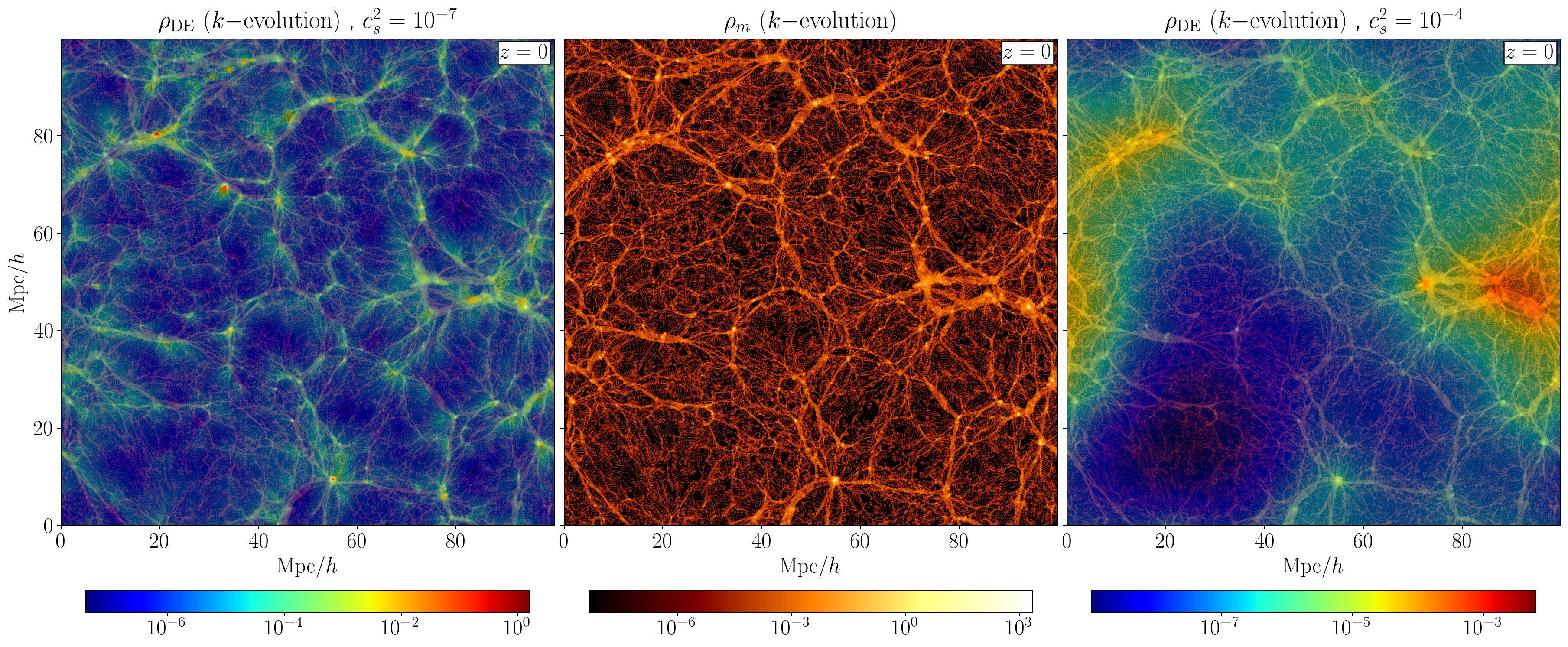}
    \caption{Snapshots displaying matter density (middle panel), dark energy density having a squared sound speed of $c_s^2 = 10^{-7}$ overlaid on matter density (left panel), and dark energy density with a sound speed squared of $c_s^2 = 10^{-4}$ overlaid on matter density (right panel). For low sound speeds, dark energy is more prone to the gravitational pull of matter, causing it to cluster in areas of high density. In contrast, when dark energy has a higher sound speed, its fluctuations spread quickly and it evades the confining influence of matter potential. This rapid dispersion prevents dark energy clustering which leads to less pronounced dark energy presence in dense matter regions compared to when the sound speed is low.}
    \label{fig:snapshots}
\end{figure*}

So far, the discussion has been about the upper panels of Fig.\ \ref{fig:mu_comparison}, with the underlying assumption that dark energy and dark matter maintain a full correlation. But as mentioned previously, the value of speed of sound can affect this correlation. In Fig.\ \ref{fig:corr_fact}, we show the cross-correlation coefficient, $f_\times$, for different values of the sound speed, namely $c_s^2 = 10^{-4}$, $c_s^2 = 10^{-7}$ and $c_s^2 = 10^{-9}$. At large scales, where both matter and dark energy perturbations evolve linearly, a complete correlation is maintained between them. This full correlation continues at smaller scales for $c_s^2 = 10^{-4}$, where dark energy does not cluster strongly and tightly tracks the  matter density. This is why in the lower right panel of Fig.\ \ref{fig:mu_comparison}, in both cases of $f_\times(k,z)\leq 1$ and $f_\times(k,z)=1$, the $\mu$ function is the same across all scales and redshifts.
Conversely, for smaller speeds of sound, $c_s^2 = 10^{-7}$ and $c_s^2 = 10^{-9}$, due to the reasons outlined earlier regarding dark energy's sound horizon being within the region where matter clusters non-linearly, dark energy is able to cluster strongly and initiate its own dynamics to some extent. Due to this, as it can be seen from Fig.\ \ref{fig:corr_fact}, the moralized cross-correlation factor start to decay by moving tho small scales. Moreover, the same logic clarifies why in the lower left panel of Fig.\ \ref{fig:mu_comparison}, the $\mu$ function for the case of $f_\times(k,z)\leq 1$ undershoots the one that assumes $f_\times = 1$, converging towards $\Lambda$CDM at slightly larger scales. \\
For illustrative purposes, in Fig. \ref{fig:snapshots} we have shown 2D snapshots of matter and dark energy densities obtained from \texttt{k-evolution} with a relatively high resolution; we simulated the evolution of $1200^3$ particles inside a comoving box with a side length of $100~\text{Mpc}/h$, resulting in a spatial resolution of $0.083 ~\text{Mpc}/ h$ . When the speed of sound is small, it allows dark energy to be influenced more significantly by the gravitational potentials of matter, leading it to cluster in high density regions. Conversely, for dark energy with a high speed of sound, the rapid propagation of its perturbations does not allow it to get trapped by the matter potential in the same way. This high propagation speed essentially prevent dark energy from clustering.  As a result, regions with dense matter do not see the same accumulation or influence of dark energy as they do when the speed of sound is low. 

Based on these discussions, we can conclude that the \texttt{k-evolution} results are primarily important for future tests of $k$-essence dark energy featuring small values of sound speed, an aspect that we will take into account in the construction of the emulator.
We emulate two versions of $\mu$; 1) when we assume a full  correlation between dark matter and dark energy, i.e. $f_\times = 1$ in Eq.\ \eqref{eq:mu-corr} and 2) when a realistic correlation between dark matter and dark energy is considered, meaning $f_\times \leq 1$. However, we will only discuss the results and methods for the version where $f_\times = 1$ is assumed across all scales and redshifts. This is because we use a similar approach to develop an emulator for the other version. We revisit the differences between the two in Section \ref{sec:reproduce}, where we explore how to use the emulator to reconstruct the power spectrum of the gravitational potential $\Phi$ from a matter power spectrum.

\section{Emulator Construction: Principal Stages}
\label{sec:EmuStages}
In this section, we give an overview of the main steps involved in the construction and configuration of a PCE-based emulator. It should be emphasized that the approach we use in our work is significantly influenced by the methods used in the \texttt{EuclidEmulator} papers \citep{Euclid:2018mlb,Euclid:2020rfv}, given their demonstrated efficiency.

As previously mentioned, we select $\mu$ function to be our emulation objective, and we define $\epsilon_\text{emu}$ as the general emulation error
\begin{equation}
    \epsilon_\text{emu}(k,z;{\bm p}) = \frac{\mu_\text{emulated}(k,z;{\bm p})-\mu_\text{simulated}(k,z;{\bm p}) }{\mu_\text{simulated}(k,z;{\bm p})},
    \label{eq:emu-error}
\end{equation}
where $k$ and $z$ stand for wavenumber and redshift, and ${\bm p}$ represents a set of cosmological parameter that are used in the computation of $\mu$ function. Note that in the context of \texttt{EuclidEmulator}, the approach taken involves emulating the \textit{boost} to the matter power spectrum as opposed to directly emulating the matter power spectrum itself. In our case, even though we have the option to emulate the boost of $\mu$ function by separately calculating $\mu$ from \texttt{k-evolution} and  \texttt{CLASS}, we have chosen to directly emulate $\mu$. This choice has been validated by the minimal emulation errors encountered in our case and eliminates the requirement for employing a Boltzmann solver to obtain $\mu$ afterward.

The main goal is to minimize $\epsilon_\text{emu}$, a crucial quantity that determines the accuracy of the final emulator and guides the selection of certain fixed hyperparameter values involved in the construction of the emulator. This, however, needs to be achieved while ensuring that the expenses of building the emulator remain within reasonable limits. 

The key stages in the construction of an emulator using the polynomial chaos expansion (PCE) approach can be outlined as follow:

\begin{enumerate}[label={\Roman*)},left=0pt]
    \item Sampling cosmological parameters within prior ranges for the purpose of training the emulator.
    \item Running  simulations to compute the $\mu$ function for each sampled parameter set, and storing these results in a matrix denoted as $\mathcal{D}$.
    \item Applying dimensionality reduction to the matrix $\mathcal{D}$ using principal component analysis (PCA).
    \item Creating surrogate models to represent the eigenvalues of the principal components as functions of the input cosmological parameters, through the use of SPCE.
    \item Merging the emulated eigenvalues (surrogate models) with their corresponding principal components.
\end{enumerate} 
 In order to carry out a comprehensive investigation with the goal of determining the optimal number of training samples $(n_{\text{ED}})$\footnote{In the domain of uncertainty quantification, the term experimental design (ED) refers to the ensemble of training input parameters $(X_{\text{ED}})$ and the resulting outputs $(Y_{\text{ED}})$, which in our case are generated during the first and second step of the emulator construction.} for the final simulation-based emulator to achieve a desired accuracy, and also to  gain an in-depth insight into the emulation mechanism—including its functionality, sensitivity to the parameter space’s dimensionality, and process of hyperparameter optimization— it is practical to start with the Boltzmann code rather than using $N$-body simulations. Hence, we begin by constructing mock emulators, following the outlined five steps, while opting to use the \texttt{Halofit} extension of \texttt{CLASS}. Subsequently, we will apply our findings from this initial phase to the final simulation-based emulator using \texttt{k-evolution}.


\subsection{Step I: Prior ranges and sampling strategy}
\label{step1} 
\subsubsection{Cosmological priors}
We define the emulator on top of a seven dimensional cosmological parameters space, encompassing \{$\Omega_\text{b}$, $\Omega_{\text{cdm}}$, $n_s$, $h$, $w_0$, $\log c_s^2$ , $A_s$\}. To ensure consistency with \texttt{EuclidEmulator}2, for each parameter we have assumed uniform priors within ranges that are closely aligned with those  in \texttt{EuclidEmulator}2, as shown in Table \ref{tab:ParameterBox}.
\begin{table}
\centering
\caption{Parameter ranges for the emulator.}
\begin{tabular}{|c|c|}
\hline
Parameter & Range \\
\hline\hline
$\Omega_\text{b}$ & [0.04, 0.06] \\\hline
$\Omega_{\text{cdm}}$ & [0.20, 0.34] \\\hline
$n_s$ & [0.92, 1] \\\hline
$h$ & [0.61, 0.73] \\\hline
$w_0$ & [-1.3, -0.7] \\\hline
$\log c_s^2$ & [-10, -5] \\\hline
$A_s$ & [$1.7 \times 10^{-9}, 2.5 \times 10^{-9}$] \\
\hline
\end{tabular}
\label{tab:ParameterBox}
\end{table}
In the process of training the emulator, each cosmology we sample from the parameter box assumes a fixed value  for $\Omega_\text{rad}$, corresponding to the CMB temperature $T_\text{CMB} = 2.725$~K. Meanwhile, the dark energy density parameter $\Omega_\text{DE}$, is calculated using the flatness condition $\Omega_\text{tot} = 1$, expressed as:
\begin{equation}
    \Omega_\text{tot} = \Omega_\text{DE} +  \Omega_\text{cdm} + \Omega_\text{b} + \Omega_\text{rad}.
\end{equation}
Differently to the case of \texttt{EuclidEmulator}2, where a dynamical dark energy model is adopted using both $w_0$ and $w_a$, our parameterisation for the $k$-essence model of dark energy employs $w_0$ and the speed of sound, $c_s$.
For all cosmological models considered in this study, we set $w_a$ to zero, resulting in a more simplified yet robust description of the dark energy.
Furthermore we have limited our analysis to small values of the sound speed between $c_s^2 = 10^{-10}$ and $c_s^2 = 10^{-5}$ since as showed in the previous section larger values yield results that are consistent with those obtained from Boltzmann code.

\subsubsection{Sampling strategy}
The effectiveness of an emulator depends significantly on the strategy used for sampling from the parameter space. As a widely adopted and popular option, we choose latin hypercube sampling (LHS) \citep[ e.g.~][]{McKay1979} for several reasons: 

\begin{enumerate}[label=(\roman*), leftmargin=*]
    \item LHS ensures that each parameter is uniformly sampled across its entire range. This uniformity is crucial in high-dimensional spaces where the emulator needs to capture the wide variability of parameters.
    \item Compared to simple random sampling, LHS can provide more extensive coverage of the parameter space with fewer samples. This efficiency is especially valuable in context of expensive simulations, where the number of training simulations is restricted by computational resources.
    \item The stratified sampling approach of LHS contributes to better predictive accuracy of the emulator. It allows for a more representative and diverse set of simulation runs.
\end{enumerate}

We produce Latin hypercube samples of varying sizes, $n_\text{ED} = \{10, 25, 50, 100, 125, 200, 300\}$, to subsequently identify the optimal sample size that meets our required accuracy standards. It is important to note that before initiating the LHS process, the total number of samples has to be specified. LHS inherently leads to non-unique outcomes due to the random nature of the sampling within each stratum of the parameter space. To address this non-uniqueness in order to optimize our sampling strategy, we employ the method proposed in \cite{Euclid:2018mlb,Euclid:2020rfv}; For each predetermined sample size, we generate $10^4$ distinct LHS sets; the optimal set is then chosen based on the criterion of maximizing the minimal Euclidean distance between points, ensuring a sample distribution that is as spread out and evenly distributed as possible.


\subsection{Step II: Computation of the {\boldmath{$\mu$}} function}
\label{sec:comput_mu}
Following the sampling of cosmological parameter sets, for each set we run simulations (\texttt{Halofit} in the case of mock emulator) to calculate the $\mu$ function. The results of this computation are then stored in a dataset denoted as $\mathcal{D}$. This dataset is represented as matrix with $n_\text{ED}$ rows and $n_k . n_z$ columns
\begin{align*}
\mathcal{D} &= \\ 
& \scalemath{0.72}{
        \begin{bmatrix} 
        \mu_1(k_1,z_1) & \dots & \mu_1(k_{n_k},z_1) & \mid & \dots &\mid & \mu_1(k_1,z_{n_z}) & \dots & \mu_1(k_{n_k},z_{n_z}) \\
        \mu_2(k_1,z_1) & \dots & \mu_2(k_{n_k},z_1) & \mid & \dots &\mid & \mu_2(k_1,z_{n_z}) & \dots & \mu_2(k_{n_k},z_{n_z}) \\
        \mu_3(k_1,z_1) & \dots & \mu_3(k_{n_k},z_1) & \mid & \dots &\mid & \mu_3(k_1,z_{n_z}) & \dots & \mu_3(k_{n_k},z_{n_z}) \\
        \vdots & \vdots & \vdots & \mid & \dots &\mid& \vdots & \vdots & \vdots \\
        \mu_{n_\text{ED}}(k_1,z_1) & \dots & \mu_{n_\text{ED}}(k_{n_k},z_1) & \mid & \dots &\mid& \mu_{n_\text{ED}}(k_1,z_{n_z}) & \dots & \mu_{n_\text{ED}}(k_{n_k},z_{n_z}) \\ 
        \end{bmatrix},}
\end{align*}   

It is important to note that redshift is not included in the cosmological parameters that are sampled. Instead, the matrix $\mathcal{D}$ stores $\mu$ functions at specific redshifts corresponding to various interval steps of the simulation\footnote{In fact, similar to \cite{Euclid:2018mlb,Euclid:2020rfv}, we store the logarithms of these $\mu$ functions inside the matrix $\mathcal{D}$ as doing so will result in lower emulation error in the end.}. Consequently, for any randomly chosen cosmological parameter from Table \ref{tab:ParameterBox}, the final emulator produces the $\mu$ function at redshifts and wavenumbers for which it is trained. To enable the user to obtain the emulated $\mu$ values at redshifts not present in the matrix, we implement linear interpolation between the two nearest columns that encompass the requested redshift. Since for now, our current focus is solely on studying the emulation error, we compute the spectra at only five specific redshifts: $\{0, 0.5, 1, 2, 3\}$, across a range of $1000$ wavenumbers within $10^{-2} h \, \text{Mpc}^{-1} < k < 10 h \, \text{Mpc}^{-1}$.


\subsection{Step III: Principal component analysis (PCA) }
Given the high dimensionality of the matrix $\mathcal{D}$, computational processing is notably challenging. Furthermore, not all details within the data, such as simulation noises and other non-physical phenomena, are desirable for our analysis and are thus intended to be removed. In this context, techniques for reducing dimensionality, such as PCA, are highly efficient. An additional compelling reason for employing PCA is the limitations of our emulation method. A basic PCE is inherently designed to predict scalar outputs and can not be directly applied to non-scalar quantities like the $\mu$ function. By using PCA, we can  decompose the covariance matrix of the dataset $\mathcal{D}$ to identify the principal components and extract their associated eigenvalues. These eigenvalues being scalar quantities, can then be emulated. 

PCA decomposes a dataset into orthogonal principal components, where each component corresponds to an axis along which the variance of the projected data is maximized. The first axis, known as the first principal component ($\text{PC}_1$), captures the largest proportion of the dataset's variance. Subsequently, the second principal component ($\text{PC}_2$) accounts for the maximum variance residual from that captured by $\text{PC}_1$, under the constraint of orthogonality to $\text{PC}_1$. This process iterates across all dimensions of the dataset, allowing for the identification of up to $n$ principal components within an $n$-dimensional space, each orthogonal to the others and sequentially optimizing the variance accounted for. For additional information on PCA, see e.g Chapter 8 of \cite{geron2022hands}.

To determine the principal components of the data matrix $\mathcal{D}$, we employ a matrix factorization method called the Singular Value Decomposition (SVD) technique. Moreover, PCA presupposes that the dataset is centered; thus, prior to applying SVD, the dataset $\mathcal{D}$ is centered by subtracting the mean of each column from its respective elements. This preprocessing step results in a centered dataset $\mathcal{D}_{\text{centered}}$, which SVD then decomposes into three matrices as shown below

\begin{equation}
\begin{aligned}
\mathcal{D_{\text{centered}}} &= \mathbf{U}~ \mathbf{\Sigma}~ \mathbf{V^T}\\
& =\sum_{i=1}^{n_{\mathrm{ED}}} \lambda_{i}
(\Omega_b, \Omega_\text{CDM}, n_s, h,w_0, \log c_s^2,A_s)  
\mathrm{PC}_{i}(k,z).
\label{eq:svd}
\end{aligned}
\end{equation}

In this decomposition, $\mathbf{V^T}$ contains the principal components. The singular values in $\mathbf{\Sigma}$, when multiplied by  $\mathbf{U}$, correspond to the weights (or magnitudes) of these principal components, reflecting their contribution to capturing the variance in $\mathcal{D_{\text{centered}}}$.
While there could potentially be $n_k . n_z$ principal components for $\mathcal{D}_\text{centered}$, the summation effectively considers only those up to $n_{\mathrm{ED}}$, as the eigenvalues $\lambda_i$ for $i > n_{\mathrm{ED}}$ vanish. 

Once all the principal components are determined, the matrix $\mathcal{D}_\text{centered}$ can be projected into a $d$-dimensional space. This is achieved by considering only the first $d$ columns of $\mathbf{V}$, which correspond to the first $d$ principal components. But, rather than selecting the number of principal components arbitrarily, a preferable strategy is to initially determine $a_\text{PCA}$, the total variance we wish to retain in the projection, and then accordingly choose the appropriate number of dimensions, $d$, to which we reduce the dimensionality. Following this approach, the inverse transformation allows for the reconstruction of $\mathcal{D_{\text{centered}}}$ back into its original dimensional space, as shown in Eq.\ \eqref{eq:svd}. However, due to the loss of information during the projection, this will not be an exact match to $\mathcal{D_{\text{centered}}}$ (unless all the principal components are considered in the summation). The reconstruction error can be measured by computing the mean squared distance between the two datasets, i.e. between $\mathcal{D_{\text{centered}}}$ and its reconstructed counterpart.

We generate a sequence of $a_{\text{PCA}}$ values, including $a_{\text{PCA}} = \{0.9, 0.99, \ldots, 0.999999\}$, for the purpose of determining the most suitable value through a grid search, which will be explained in detail later.


\subsection{Step IV: Polynomial chaos expansion (PCE) }

Emulating the scalar outputs $\lambda_i$, obtained from the PCA in the previous step, constitutes the essential phase of this work. Since each $\lambda_i$ captures a distinct aspect of variance in the data, it is crucial to emulate them separately. To do this, we employ PCE method integrated within the \texttt{UQLab}\footnote{\url{https://www.uqlab.com}} software to provide a functional approximation of each eigenvalue. PCE achieves this by representing the model's spectral characteristics on an appropriately constructed orthonormal basis of polynomial functions. In our case, the expansion is expressed as follows

\begin{equation}\label{PCE}
\lambda_{i}\left(\Omega_b, \Omega_\text{CDM}, n_s, h,w_0, \log c_s^2,A_s\right) 
  = \sum_{\bm\alpha \in \mathbb{N}^M}\beta_{\bm\alpha}\Psi_{\bm\alpha}({\bf X}).
\end{equation} 
Here, $\bm\alpha = \{\alpha_1,\alpha_2,...,\alpha_M\}$ represents a multi-index, $\Psi_{\bm\alpha}$ corresponds to the orthonormal multivariate polynomial basis, and $\beta_{\bm\alpha}$ is its associated coefficient. Moreover, ${\bf X} = \{X_1,X_2,...,X_M \}^\text{T}$ denotes the input random vector of the cosmological parameters, and $M$ indicates the dimensionality of the parameter space, which in our case is seven.
The multivariate polynomials $\Psi_{\bm\alpha}$, are constructed through the product of univariate polynomials $\phi_{\bm\alpha}^{(j)}(X_j)$, with the specific type of each univariate polynomial determined by the distribution of the respective input variables $X_i$.
Given the uniform distribution of input cosmologies in our context, each constructed $\phi_{\bm\alpha}^{(j)}(X_j)$ is a member of the Legendre polynomial family,
\begin{equation}
    \Psi_{\bm\alpha}(\bm X) \equiv \prod_{j=1}^{M}\phi_{\alpha}^{(j)}(X_j) = \prod_{l=1}^M \sqrt{2\alpha_l+1} P_{\alpha_l}(X_l).
\end{equation}

\subsubsection{Sparse polynomial chaos expansion: Truncation Strategy}
For practical implementation of the Eq.\ \eqref{PCE}, the summation has to be truncated to fit within the computational limitations. A common strategy to achieve this involves restricting the number of multi-indices which results in the reduction of the number of multivariate polynomial bases $\Psi_{\bm\alpha}$ participating in the summation. This restriction is implemented by constructing a limited index set (also known as the truncation set) denoted as $\mathcal{A}$, such that $\mathcal{A} \subset \mathbb{N}^M $.

 A logical initial step in constructing  $\mathcal{A}$ would be to filter the polynomial bases in order to keep only those whose total degree does not exceed $p$. This leads to the definition of an index set characterized as follow:
\begin{equation}
 \mathcal{A}^{M,p} = \{ \bm\alpha \in \mathbb{N}^M, ||\bm\alpha||_1 \leq p\},
 \label{indexp}
\end{equation}
 where the term $||\bm \alpha||_1$ represents the length of the vector $\bm \alpha$, and is defined as
\begin{equation}
  ||\bm \alpha||_1 = \sum_{j=1}^{M}\alpha_j .
  \label{length}
\end{equation}
However, this truncation scheme solely may not sufficiently reduce the complexity in practical situations, since the total number of polynomial terms, denoted by $N$, follows the formula
\begin{equation}\label{NumTerm}
N = 
\begin{pmatrix} 
p + M \\ 
p 
\end{pmatrix} 
= \frac{(p + M)!}{p! \, M!} .
\end{equation}
Consequently, the number of terms grow rapidly with the
increase in the number of input variables, making the process computationally challenging and often impractical in scenarios involving high-dimensional parameter spaces.

To address this, we turn to strategies rooted in the \textit{sparsity-of-effects} principle. This principle is based on the observation that in many complex systems, only a few interaction terms (low-order interactions) significantly influence the behaviour of systems. Most of the higher-order interactions or terms have negligible effects. Therefore, identifying and focusing on these lower order influential terms can lead to a more efficient and computationally manageable model. For that, we employ two widely used truncation scheme as detailed in \cite{BlatmanThesis,BLATMAN20112345, PCEManual}.

One such approach is known as \textit{hyperbolic truncation}. This strategy modifies (extends) the standard degree-based truncation scheme to maintain significant lower-order terms, further penalizing higher-rank indices.
 Unlike the standard method, which includes all terms up to a specified degree, hyperbolic truncation concentrates on terms with a lower combined total degree, determined by the $q$-norms condition. This corresponds to an index set defined as
\begin{equation}
  \mathcal{A}^{M,p,q} = \{ \bm\alpha \in \mathcal{A}^{M,p},||\bm\alpha||_q \leq p\},
\end{equation} 
where
\begin{equation}
 ||\bm\alpha||_q = \left (\sum_{j=1}^{M} {\alpha_j}^q \right)^{1/q} ~~~,~~~ 0< q \leq 1 .
 \label{qnorm}
\end{equation}
Conceptually, the hyperbolic condition can be seen as the selection of polynomial terms that fall under a hyperbolic curve within the space defined by the multi-index $\bm\alpha$ \citep[see Fig.\ 1 of][]{PCEManual}.

Another complementary strategy is the \textit{maximal interaction}. This technique limits the highest order of interactions between variables in the constructed polynomials, rather than just limiting the degree of individual polynomials. The model can be specifically adjusted to limit the number of non-zero elements in the multi-index $\bm\alpha$ to a maximum value $r$, where $r \leq M$. This is mathematically represented as

\begin{equation}
  \mathcal{A}^{M,p,r} = \{ \bm\alpha \in \mathcal{A}^{M,p}, ||\bm\alpha||_0 \leq r\} ,
\end{equation}
with
\begin{equation}
    ||\bm \alpha||_0 = \sum_{j=1}^{M} 1_{\{\alpha_j > 0\}} ,
\end{equation}
being defined as the rank of $\bm\alpha$.
In practical terms, this translates to allowing only $r$ or fewer univariate polynomials $\phi_{\alpha}^{(i)}(X_i)$ to participate in constructing each term of the multivariate polynomial bases. Essentially, it restricts the model to consider interactions among at most $r$ input parameters in each $\Psi_{\bm\alpha}$.

Integrating both truncation methods, the definitive index set can be formulated as
 \begin{equation}
  \mathcal{A}^{M,p,q,r} = \{ \bm\alpha \in \mathcal{A}^{M,p},||\bm\alpha||_q \leq p~~ {\text{and}}~~ |
  |\bm\alpha||_0 \leq r  \}.
  \label{finite}
 \end{equation}
Thus the SPCE of each considered eigenvalues  reads
 \begin{equation}
\lambda_{i}  
  \approx \sum_{\bm \alpha \in \mathcal{A}_i^{M,p,q,r}}\beta_{\bm\alpha}\Psi_{\bm\alpha}({\bf X}).
  \label{finpce}
\end{equation} 

\subsubsection{Calculation of the coefficients}
In \citet{BlatmanThesis}, several methods for calculating the expansion coefficients $\beta_\alpha$ are proposed. However, \texttt{UQLab} implements only those methods that involve post-processing of model evaluations. The method we employ here is the Least Angle Regression (LARS) technique \citep{Efron2004}. This approach solves the following optimization problem

\begin{equation}
\hat{\boldsymbol{\beta}}_{i, \alpha}=\underset{\boldsymbol{\beta}_i \in \mathbb{R}^{\left|\mathcal{A}_i^{p, q, r}\right|}}
{\operatorname{argmin~}} \mathbb{E}\left[\left(\boldsymbol{\beta}_i^{\top}
\Psi_i(\boldsymbol{X})-\lambda_{i,\text{true}}(\boldsymbol{X})\right)^2 +\gamma\|\boldsymbol{\beta}\|_1\right].
\label{lars}
\end{equation}  
Here, the loss function is defined as the mean squared error between the emulated and true eigenvalue of each principal component. Additionally, it includes a regularization term
\begin{equation}
\gamma \|\boldsymbol{\beta}\|_1 = \gamma\sum\limits_{\alpha \in \mathcal{A}_i^{p, q, r}}|\beta_{i,\alpha}| ,
\end{equation}
which is introduced to penalize high-rank solutions in the minimization process. Further details on the calculation of PCE coefficients based on LARS are available in \cite{PCEManual,Efron2004}.

\subsubsection{Basis adaptivity of PCE}
Determining the most suitable finite basis for an effective PCE modeling is often demanding in practical contexts. The basis adaptivity algorithm in PCE addresses this challenge by selecting the most relevant polynomial terms from a broad range of candidates. The process involves an iterative approach, where the degree $p$ and the $q$-norm parameters are adjusted to optimize the validation error of the emulator. Based on predefined set of $p$ and $q$ values, the algorithm executes a sequence of operations that can be outlined as follow:

\begin{enumerate}[label=(\roman*), leftmargin=*]
    \item Starting the process by constructing a basis with initial values $p = p_0$ and $q = q_0$.
    \item Computing the PCE coefficients and measuring the generalization error of the model. This measurement involves using the Leave-One-Out (LOO) cross-validation error, $\varepsilon_{\text{LOO}}$, which is used to prevent overfitting. It is calculated by constructing $n_\text{ED}$ separate emulated eigenvalues, $\lambda^{\backslash i}_\text{emu}$, with each being built upon a reduced experimental design $X \backslash x(i) = \{x(j), j = 1, \ldots, n_\text{ED}, j \neq i\}$, where $\backslash x(i)$ denotes the exclusion of the $i$-th data point from $X$ . The method sequentially leaves out each data point, trains the model on the remaining data (on $n_\text{ED}-1$ samples), and then assesses its accuracy by comparing the model's predictions at the excluded data point $x(i)$ with the actual value $\lambda_\text{true}$. The LOO error is thus given by
\begin{equation}
\varepsilon_{\text{LOO}} = \frac{\displaystyle\sum\limits_{i=1}^{n_\text{ED}} \left(\lambda_\text{true}(x(i)) - \lambda^{\backslash i}_{\text{emu} }(x(i))\right)^2}{\displaystyle\sum\limits_{i=1}^{n_\text{ED}} \left(\lambda_\text{true}(x(i)) - \hat{\mu}_\lambda\right)^2} ,
\end{equation}
where the denominator serves to normalize the error by scaling it against the variance of the true output. Note that The computation of $\varepsilon_{\text{LOO}}$ is carried out entirely within the experimental design, which includes both the training dataset and the corresponding responses, and does not involve a separate test set.

    \item Evaluating the calculated error against a predetermined threshold. If the error is within this threshold or if the algorithm reaches its maximum iteration limit, it terminates and selects the PCE model with the minimal $\varepsilon_{\text{LOO}}$. If not, the algorithm adjusts the parameters by incrementing $p$ or $q$, and restarts the process from the basis generation phase.
\end{enumerate}

The default setting in \texttt{UQLab} is such that if $\epsilon_\text{LOO}$ does not decrease over two consecutive iterations, the degree-adaptive scheme stops increasing the maximum degree. For the case of q-norm adaptivity, when the polynomial degree is low, the effect of the q-norm is minimal and usually does not lead to the inclusion of extra basis functions, resulting in the $\epsilon_\text{LOO}$ remaining unchanged. Because of that, in \texttt{UQLab}, iterations of $q$-norm are only considered \textit{relevant} if the PCE's basis size or the $\epsilon_\text{LOO}$ are affected. Subsequently, the increase of $q$-norm is automatically stopped if there is no decrease in the $\epsilon_\text{LOO}$ over two \textit{relevant} iterations. In our configuration, however, we deactivated this early stop mechanism and allowed for a complete examination of all $q$-norm candidates for every degree. Doing so, we ensured that the optimal candidate is chosen after all possibilities have been considered.

\subsection{Step V: Combining emulated {\boldmath$\lambda_i$} and principal components}
In the last phase, the emulated $\lambda_i$ are merged with the corresponding principal components that we have taken into account in the SVD process. It is important to recall that we centered the $\mathcal{D}$ matrix before conducting SVD by subtracting the mean of each column from the respective $\mu$ function.  Therefore, to formulate the emulated matrix we must re-add these mean values to the combined set of emulated eigenvalues and principal components. This leads to the final derivation of the emulated $\mu$ as shown in the following equation
\begin{equation}
 \mu_\text{emu}(k, z ; \boldsymbol{p}) \approx \mu_{\mathrm{PCA}}(k, z)+\sum_{i=1}^{n_{\mathrm{PCA}}}  
 \sum_{\alpha \in \mathcal{A}_i^{p, q, r}} \hat{\beta}_{i, \alpha} \Psi_{i}^{\alpha}
 (\boldsymbol{p}) \mathrm{PC}_{i}(k, z).
\end{equation} 

\section{Fine-Tuning emulator's parameters}
\label{sec:fine-tuning}
In the preceding section, we have seen how the construction of the emulator at various stages of its assembly is fundamentally influenced by a series of critical parameters, namely the Latin hypercube sample size ($n_{\text{ED}}$), the percentage of the dataset's variance we wish to retain during PCA ($a_{\text{PCA}}$), the maximum total degree of the polynomial bases ($p$), and the truncation parameters ($q$ and $r$). So, it becomes clear that choosing the best values for these parameters is crucial for ensuring that the emulator operates at its highest level of performance. To this end, we employ a grid search analysis.
In a grid search we explore in details a manually specified subset of these  parameters\footnote{Here, there is no need to specify grid values for $r$; \texttt{UQLAB} automatically determines the optimal $r$ during polynomial coefficient computation. For instance, if $r$ is found to be $3$, it implies that polynomials with interactions beyond $3$ do not significantly contribute to the model, as their coefficients would be negligible or zero.}. The objective is to evaluate different combinations of parameter values to determine which combination yields the best performance in terms of the emulator's accuracy and computational efficiency. For this purpose, we have created the following subset of parameters
\begin{equation}
 \begin{aligned}
  n_{\text{ED}} &= \{ 10,25,50,100,125,200,300 \} ,\\
  a_{\text{PCA}} &= \{0.9,0.99,0.999,..., 0.999999\} ,\\ 
  p &= \{3,4,5,...,20 \} ,\\
  q &= \{0.25,0.30,0.35,...,1\} .\\
  \label{grid} 
 \end{aligned}
\end{equation}

The fine-tuning of the emulator's parameters is strategically divided into two steps. In the first step, the hyperparameters $n_{\text{ED}}$ and $a_{\text{PCA}}$ are adjusted based on their performance against a separate test set. Since the largest emulation errors are often found near the edges of the parameter spaces, we have generated $10^6$ random test samples and then filtered out $15\,000$ samples that fall inside a hypersphere of radius 1, inscribed in the normalized parameter box, where each parameter is mapped between $-1$ and $1$. This approach effectively concentrates our validation on the more reliable central region while minimizing the impact of edge-related errors. Note that in addition to the cosmological parameters listed in Table \ref{tab:ParameterBox}, the test set that we generate also includes the sum of neutrino masses in the range $0.0 ~\text{eV} \leq \sum m_\nu \leq 0.15~\text{eV}$, thus making the test sets effectively eight-dimensional.
It is important to clarify that the emulator itself is not trained to include the sum of neutrino masses as a variable. However, as we will discuss later, the inclusion of this parameter has a minimal impact on the $\mu$ function, indicating that the emulator is not particularly sensitive to it. By adding this parameter to the test set, users are given the option to include the sum of neutrino masses in their analyses for a range of applications. However, it is worth mentioning that the effect of including the sum of neutrino masses is only introduced through the change in either $\Omega_{\rm m}$ or $\Omega_{\rm DE}$, as massive neutrinos contribute to the total energy density with $\Omega_{\rm tot}=1$.

In the second step, the remaining parameters, $p$, $q$, and $r$, are fine-tuned internally using the LOO error measurement, which, as discussed before, evaluates the cross-validation error exclusively based on the experimental design, i.e. the training datasets and their corresponding outputs.

\begin{figure}
    \centering
    \includegraphics[width=0.45\textwidth]{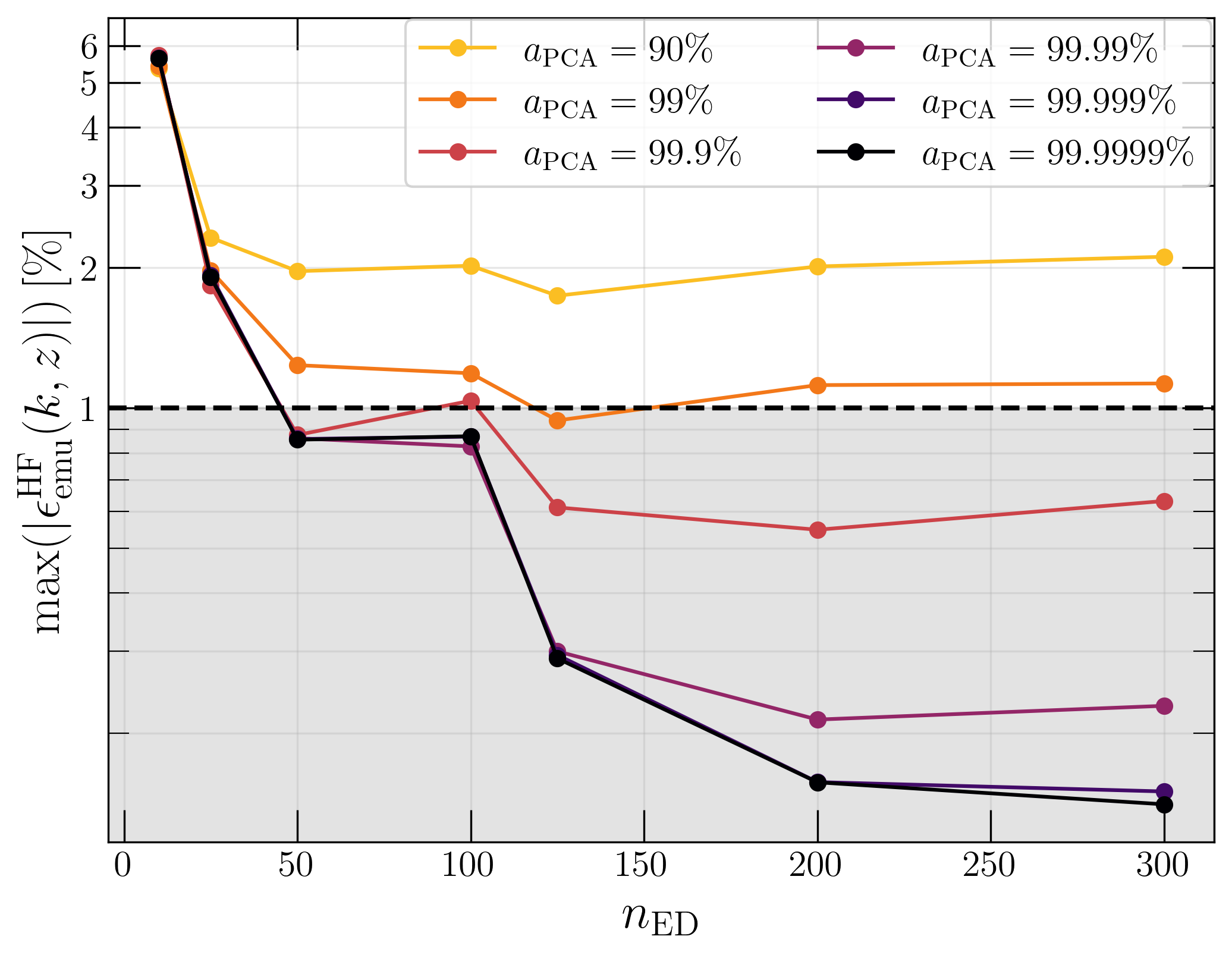}
    \caption{Demonstration of the maximum emulation error (y-axis) as a function of the number of training samples, $n_{\text{ED}}$ (x-axis), and the retained fractional variance, $a_{\text{PCA}}$, indicated by different line colors. The graph shows the dual impact of the training sample size and the variance retained during PCA on the accuracy of the emulator tested against $15 \times 10^3$ test samples. For each set of $n_\text{ED}$ and $a_{\text{PCA}}$, the parameters $p$, $q$ and $r$ are selected based on the LOO error.}
    \label{fig:a_pca}
\end{figure}

Similar to the approach taken in \cite{Euclid:2018mlb,Euclid:2020rfv}, we conducted an analysis, illustrated in Fig.\ \ref{fig:a_pca}, to examine how the maximum emulation error varies with the number of training samples and the fractional variance retained during PCA.
\begin{figure}
    \centering
    \includegraphics[width=0.45\textwidth]{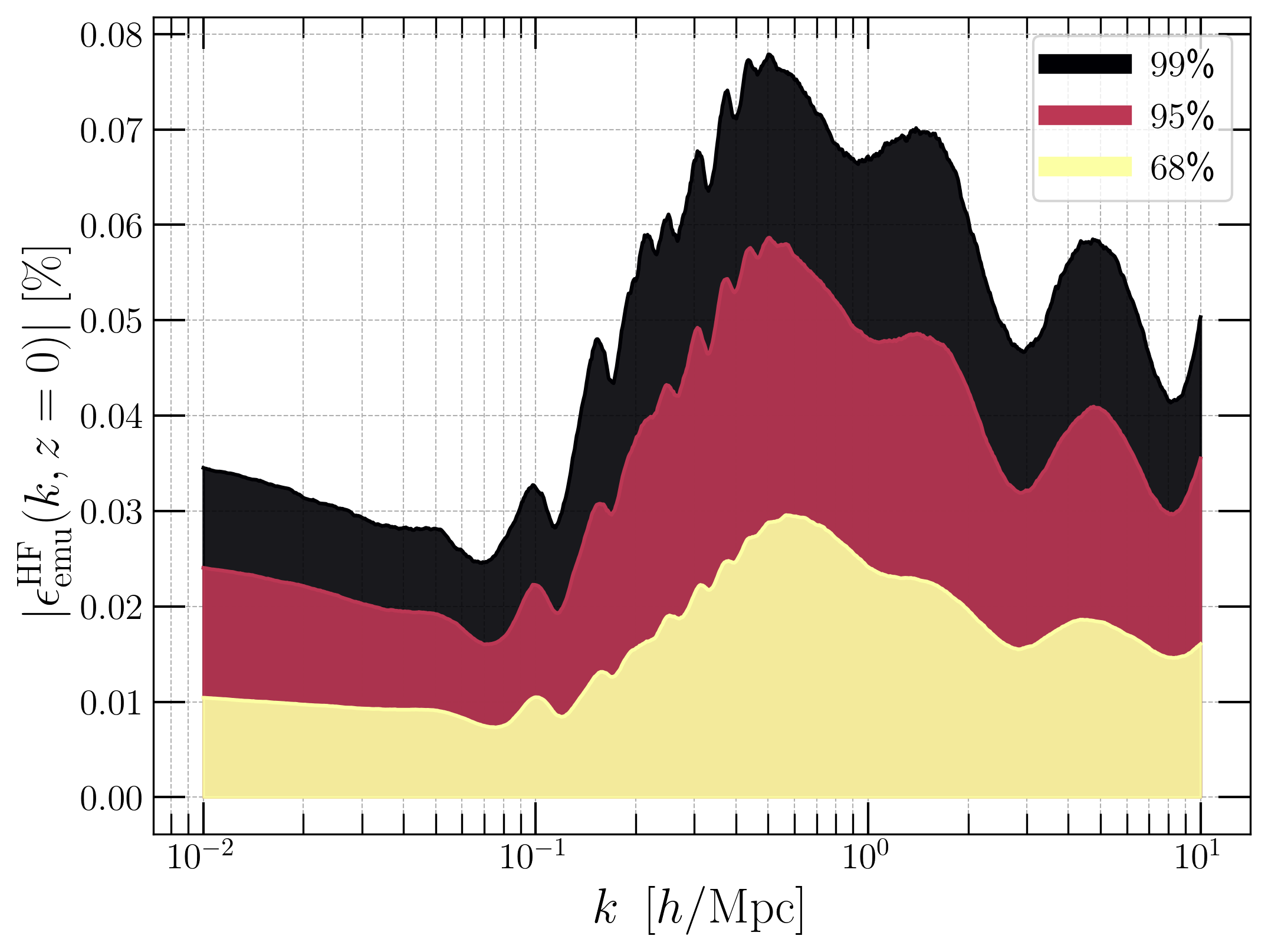}
    \caption{Emulation error distribution across $1000$ wavenumbers at $z=0$, evaluated on $15 \times 10^3$ eight-dimensional test sets, at the 99th percentile (black area), 95th percentile (red area), and 68th percentile (yellow area). These percentiles represent the values below which $99\%$, $95\%$, and $68\%$ of data points, respectively, are included.}
    \label{fig:percentile}
\end{figure}
In Fig.\ \ref{fig:a_pca}, our analysis shows that even when up to $90\%$ of the total variance is preserved in PCA, the maximum emulation error does not fall below $1\%$, despite using as many as $300$ training samples. This outcome, highlighted by the plateaus in the graph's curves for different values of $a_\text{PCA}$, implies that a mere increase in training sample size, beyond a certain threshold, is ineffective in further reducing the maximal emulation error when the retained fractional variance during PCA is insufficient.\\
Based on this analysis, we have decided to use $200$ training samples in  the construction of the \texttt{k-evolution}-based emulator, ensuring that $a_\text{PCA} = 99.999\%$ of the total variance is preserved through PCA. In the case of the mock emulator, this amount of $a_\text{PCA}$ has corresponded to $11$ principal components, resulting in $11$ separate PCE for emulating the associated eigenvalues. Details regarding the \texttt{k-evolution}-based emulator will be provided in Section \ref{sec:simulation-based emulator}.

In Fig.\ \ref{fig:percentile}, the percentile representation of the mock emulator, constructed with these $11$ principal components is depicted. The accuracy was evaluated using the previously mentioned $15\,000$ test samples. We also conducted an in-depth performance analysis of the mock emulator within the mentioned eight-dimensional parameter space, constructing 28 distinct 2D planes by pairing different parameters. On each plane, we created a $40 \times 40$ grid of test sample points, resulting in a total of $44\,800$ samples across all the planes, and then tested the emulator's performance at each point while keeping the other parameters, not involved in the respective plane, constant. Doing so, we could analyze the effects of specific parameter interactions on the emulator's accuracy and study the emulator's performance across various areas of the parameter spaces. To visualize the results, we generated heat maps for each of the 28 planes, as shown in Fig.\ \ref{fig:planes}.


\section{Sensitivity analysis based on Sobol indices}
\label{sec:Sobol}
To assess the influence of individual input parameters on the eigenvalues $\lambda_i$, we use the Hoeffding-Sobol decomposition technique, as detailed in \cite{Sobol2001, Hoeffding1948}. This method employs variance-based sensitivity analysis, where Sobol indices (denoted by $S$) are used to distribute the variance of $\lambda_i$ across each input parameter, and across combinations of parameters including pairs, triplets, and higher-order interactions.
For the computation of Sobol indices, we adopt a polynomial expansion approach (PCE-based Sobol indices, as referenced in \cite{SUDRET2008964, sensitivity2022}) instead of relying on traditional Monte Carlo simulations. 
\begin{figure}
    \includegraphics[width=0.45\textwidth]{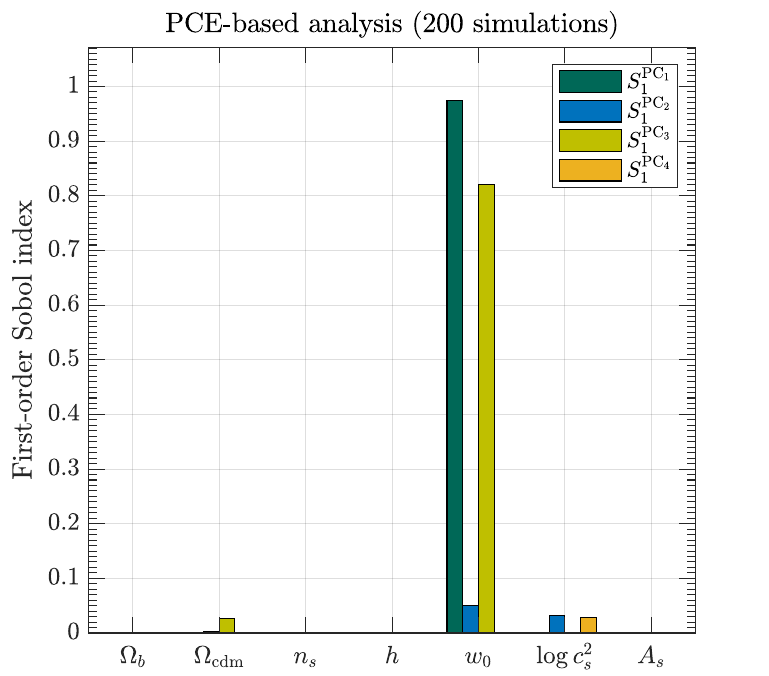}
    \caption{First-order Sobol indices for the first four principal components.}
    \label{fig:Sobol}
\end{figure}
More description of this approach, including its theoretical framework, can be found in Appendix A.

\begin{figure}
    \includegraphics[width=0.45\textwidth]{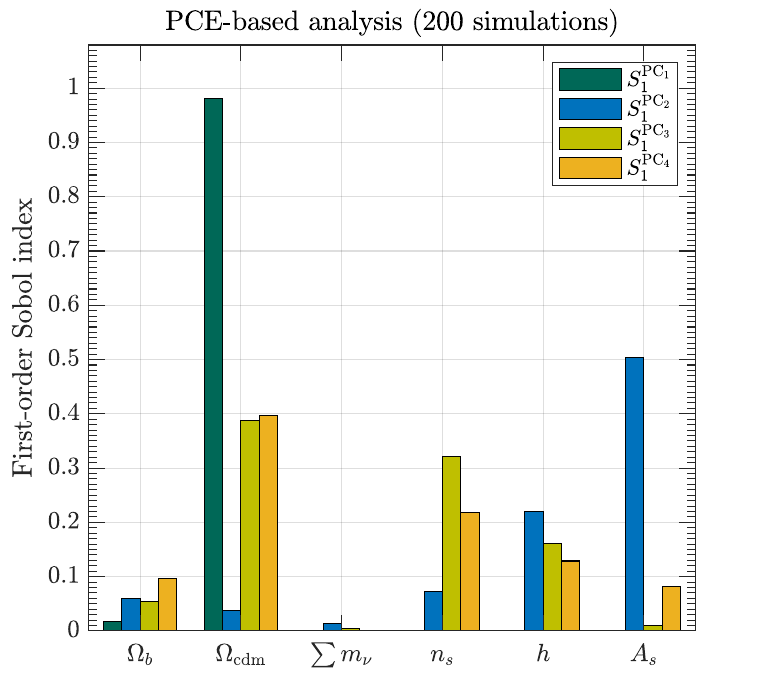}
    \caption{First-order Sobol indices for the first four principal components. $w_0$ and $c_s^2$ are maintained as fixed values (and hence not involved in the PCE) to explore the relative impacts of other parameters.}
    \label{fig:SobolNotw0cs2}
\end{figure}

Fig.\ \ref{fig:Sobol} shows the result of the first-order Sobol indices ($S_1$) for the first four principal components, based on $200$ simulations. Clearly, $w_0$ has the most impact on the output model, as it holds the highest first-order Sobol index in the first principal component. 
It is important to recognize that a significant portion of the total variance in the original data, about $97.77 \%$, is already captured in the projection onto the first principal component. Consequently, the significance of the first order Sobol indices for the cosmological parameter diminishes generally as one progresses to higher principal components. 
Following the same logic, $c_s^2$ and $\Omega_\text{cdm}$ are identified as the subsequent influential cosmological parameters due to their secondary contribution to the second and third principal components. These can be justified by considering that $w_0$ and $\Omega_\text{cdm}$ collectively influence the amplitude of the $\mu$ function, whereas $c_s^2$ affects its cut-off scale, transitioning from large to small scales.\\
While Sobol analysis effectively reveals the relative influence of cosmological parameters on the $\mu$ function, it particularly underscores the significant influence of the $k$-essence parameters, namely $w_0$ and $c_s^2$, as somewhat expected. To better investigate the impacts of other cosmological parameters, beyond those associated with $k$-essence, we keep the $k$-essence parameters constant, setting $w_0 = -0.9$ and $c_s^2 = 10^{-7}$. We then proceed by once again conducting a targeted PCE-based Sobol analysis on the remaining parameters, including the sum of neutrino masses, to investigate its impact as well. The result of this is presented in Fig.\ \ref{fig:SobolNotw0cs2}. As can be observed from the analysis, among all the cosmological parameters, the sum of the neutrino masses exhibits the lowest impact across all the first four principal components. Due to the negligible influence of the total neutrino masses, we decided to omit it from the training set. However, users have the option to enter a value ranging from $0.0 ~\text{eV}$ to $0.15 ~\text{eV}$ in the emulator's input for their own purposes (see the explanation in Section \ref{sec:fine-tuning}). This is why in Section 4, we added $\sum m_\nu$ into the test set to demonstrate explicitly that this does not lead to unacceptable emulation errors.

\section{Simulation-based emulator}
\label{sec:simulation-based emulator}

Up to this point, our main focus has been on the configuration and the construction of the mock emulator using \texttt{Halofit} within the \texttt{CLASS} framework. The main reason for this choice was its lower computational demand compared to $N$-body simulations, which made it possible to generate extensive test datasets—a task practically impossible with $N$-body simulations. In this section, we intend to apply the insights gained from developing the mock emulator to construct an emulator based on the \texttt{k-evolution} code.

\subsection{Convergence tests}
\label{sec:ConvTest}
Before proceeding with simulations to train the emulator, conducting convergence tests is imperative. These tests aim to determine the optimal simulation parameters, such as box size, number of particles, and corresponding spatial resolution, necessary to achieve an acceptable level of accuracy. In our case, we are interested in results as accurate as possible within the $k$ value range of $10^{-2} h \, \text{Mpc}^{-1} < k < 10 h \, \text{Mpc}^{-1}$.

\begin{figure}
    \includegraphics[width=0.48\textwidth]{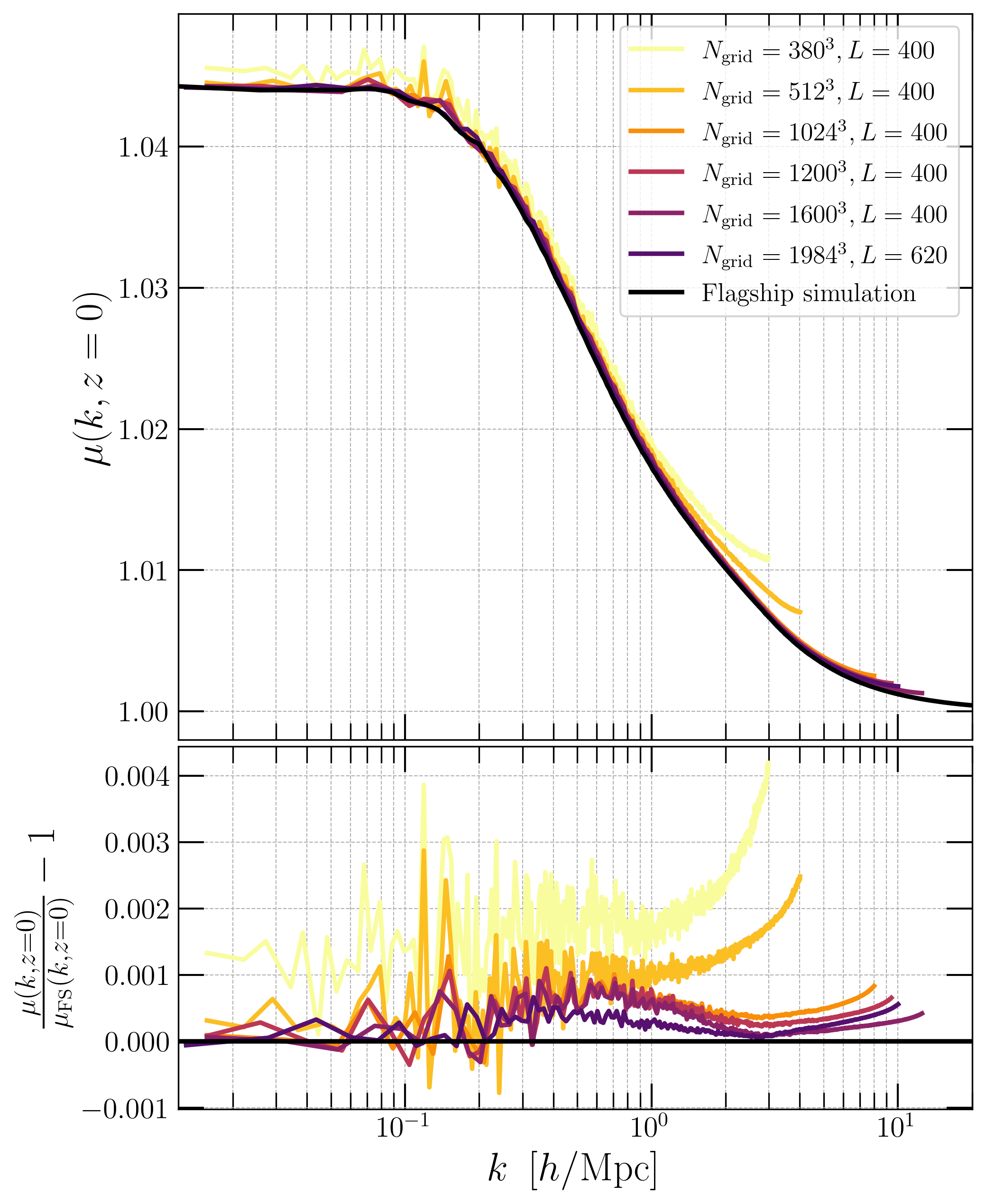}
    \caption{Convergence test analysis for the $\mu$ function, with mesh sizes of $N_\text{grid} = \{380^3, 512^3, 1024^3, 1200^3, 1600^3, 1984^3\}$ where $N_\text{pcl} = N_\text{grid}$, demonstrates that a grid size of $N_\text{grid} = 1200^3$ reaches a convergence level of $0.1\%$ within the wavenumber range of $10^{-2} h \, \text{Mpc}^{-1}$ to $10 h \, \text{Mpc}^{-1}$. Increasing the grid size to $N_\text{grid} = 1600^3$ and $N_\text{grid} = 1984^3$, results in only marginal improvement in the precision of convergence.}
    \label{fig:convergence}
\end{figure}

For the Flagship simulation, we conducted two simulations based on the cosmological parameters specified in Table \ref{table:cosmoparams}. The first simulation involved $3840^3$ particles and a box size of $L = 1280 ~h^{-1}~ \text{Mpc}$, leading to a Nyquist wavenumber of $k_\text{N} = 9.42 ~h~ \text{Mpc}^{-1}$. The second simulation contained $2304^3$ particles and a smaller box size of $L = 300 ~h^{-1} ~\text{Mpc}$, yielding a Nyquist wavenumber of $k_\text{N} = 24.12~h~ \text{Mpc}^{-1}$. By linking these two simulation we managed to provide reliable results on both large and small scales.  In our series of test simulations, we primarily focused on varying the grid size ($N_\text{grid}$) while maintaining a consistent box size for the majority of the tests. The details of these simulations are shown in Table \ref{tab:simulation_parameters}. For each test, the results were truncated at the corresponding Nyquist wavenumber to prevent sampling errors and preserve the validity of the simulation data. In Fig.\ \ref{fig:convergence} we have shown the result of these convergence tests.
\begin{table}
\centering
\caption{Simulation parameters and computational resources}
\label{tab:simulation_parameters}
\scalebox{0.85}{
\begin{tabular}{|c|c|c|c|p{1.5cm}|c}
\hline
\textbf{Label} & $N_\text{grid}$ & $L$ ($h^{-1}$ Mpc) &  $k_\text{N}$ ($h ~\text{Mpc}^{-1}$)&\textbf{Runtime (core hours)} & \textbf{Machine} \\ \hline\hline
$^1$FS  & $3840^3 $   & 1280 & 9.42 & 50688 & CSCS \\ \hline
$^2$FS & $2304^3 $ & 300 & 24.13  &   129024  & CSCS    \\\hline
Sim1  & $380^3 $  & 400 & 2.98 &     19   & Baobab        \\ \hline
Sim2  & $512^3 $  & 400 & 4.02 &     64   & Baobab        \\ \hline
Sim3 & $1024^3$   & 400  & 8.04  &      1536  & Baobab  \\ \hline
Sim4   & $1200^3 $  & 400 & 9.42 &  2560  & Baobab   \\ \hline
Sim5    & $1600^3$ & 400& 12.56  &    9627  & Baobab    \\ \hline
Sim6   & $1984^3$   & 620 & 10.05 &   17641   & Baobab  \\ \hline
\end{tabular}
}
\end{table}

Our analysis reveals that the test simulation, using $N_\text{grid} = N_\text{pcl}=1200^3$ and a box length of $L=400 h ~\text{Mpc}^{-1}$, has successfully attained a satisfactory level of convergence, falling below $0.1\%$ up to $k = 9.42 h ~\text{Mpc}^{-1}$. Further increase in number of particles and the size of the simulation volume appear to have minimal impact on improving this convergence. Given the satisfactory results achieved with the settings $ N_\text{grid} = 1200^3$ and $L = 400 h \, \text{Mpc}^{-1}$, we have decided to select these specific parameters to generate the training set for the simulation-based emulator.

We should highlight that the acceptable convergence level achieved with this relatively small number of particles, owes much to the way of calculating the $\mu$ function by dividing the square root of one power spectrum by another. This division significantly diminishes the resolution effects and the impact of cosmic variance, the latter being a consequence of the limited size of the simulation box that leads to statistical uncertainties on large scales. Such reductions in these non-physical effects are the main reason behind our decision to emulate the $\mu$ function instead of directly emulating the potential power spectrum. The high susceptibility of the potential power spectrum to such issues implies that maintaining an acceptable level of accuracy across a wide spectrum of wavenumbers would require a substantially larger number of particles and greater simulation volume. By opting for the $\mu$ function instead, we mitigate the necessity for extensive computational resources.

\begin{figure}
    \includegraphics[width=0.48\textwidth]{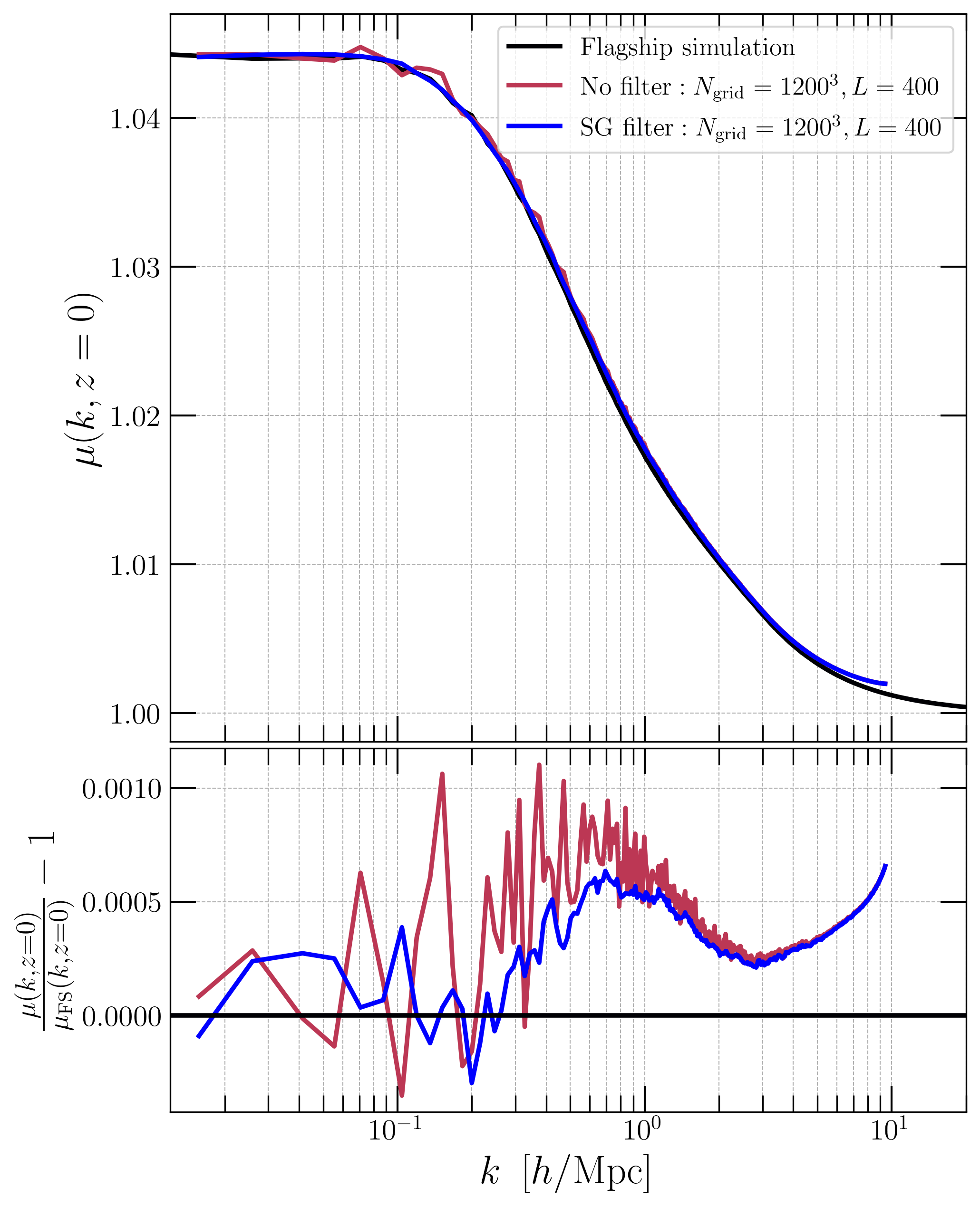}
    \caption{Demonstration of the impact of  SG filter: employing a 14-data point window and a polynomial of order 3 leads to a notable RMSE reduction, specifically in the wavenumber range from $k = 0.025 h \, \text{Mpc}^{-1}$ to $k = 1 h \, \text{Mpc}^{-1}$.}
    \label{fig:SG}
\end{figure}

 To further diminish the impact of cosmic variance on the $\mu$ function, we use the Savitzky-Golay (SG) filter \citep{Savitzky1964}. This filter essentially reduces signal noise while preserving its characteristics through the application of polynomial fits within localized segments defined by moving windows. By using an SG filter with a window length of $14$ data points and a polynomial order of $3$, i.e. fitting third-order polynomials to $14$ points at each step, we successfully achieved a notable reduction in noise across our data, especially at larger scales as depicted in Fig.\ \ref{fig:SG}.
By doing so, we could lower down the root mean square error (RMSE), described as
\begin{equation}
\text{RMSE} = \sqrt{\frac{1}{n_k} \sum_{i=1}^{n_k} \left(\frac{\mu_i - \hat{\mu}_i}{\mu_i}\right)^2} ,
\end{equation}
from $0.064\%$ to $0.043\%$ across the $k$ range of
$0.025 h\, \text{Mpc}^{-1} < k < 1 h \, \text{Mpc}^{-1}$.

\subsection{From mock to simulation-based emulator}

Following the analysis presented in Fig.\ \ref{fig:a_pca}, we proceed with the same set of $200$ Latin hypercube-sampled cosmologies for the purpose of training the \texttt{k-evolution}-based emulator.  These $200$ simulations are executed on the \textit{Baobab} and \textit{Yggdrasil} clusters\footnote{\url{https://doc.eresearch.unige.ch/hpc/start}}. As previously discussed, all simulations maintain the fixed parameters  of $N_\text{grid}=1200$ and $L=400 h \, \text{Mpc}^{-1}$, and for every cosmology set, the resulting $\mu$ function is processed through an SG filter with a window length of $14$ and a polynomial order of $3$. Here, spectra are calculated at $53$ distinct $z$-values\footnote{The selection of these redshifts is designed to ensure that the $\mu$ values across different redshifts remain as close as possible, especially at larger scales, to avoid any significant gap. Doing so helps in minimizing errors when performing linear interpolation between neighboring redshifts.} within the range of $ 0 < z < 3$  and at 1024 distinct $k$-bins. After the truncation of the results beyond the Nyquist wavenumber, this results in 591 $k$-values ranging from $10^{-2} h \, \text{Mpc}^{-1}$ to $ 9.42 h \, \text{Mpc}^{-1}$. However, prior to restoring the data in a matrix format similar to that discussed in Section \ref{sec:comput_mu}, we perform interpolation on the smoothed $\mu$ function using the available $k$-values. Subsequently, we calculate the $\mu$ values at $1000$ evenly spaced wavenumbers within the mentioned range. Therefore, the final matrix has $n_\text{ED}= 200$ rows and $n_k.n_z = 53\,000$ columns. The remaining stages of the emulator's construction are the same as explained before in Section \ref{sec:EmuStages}. This includes the principal component analysis of the matrix $\mathcal{D}$, polynomial chaos expansion of the scalar outputs $\lambda_i$, and finally re-merging the emulated eigenvalues with the corresponding principal components. Since these steps are identical to those described in the construction of the mock emulator, we will not repeat them in detail here. Instead, we will focus on presenting the  the results of these analyses.

\begin{table}
\centering
\caption{Results of the polynomial chaos expansions for \texttt{k-evolution} experimental design. The SFB (size of full basis) column indicates the total number of polynomial terms generated after truncation, using $q$ parameter, while the SSB (size of sparse basis) column reflects the number of terms with non-zero coefficients after employing the LARS method.}
\begin{tabular}{ |p{0.5cm}||p{0.5cm}|p{0.5cm}|p{0.5cm}|p{2cm}|p{0.5cm}|p{0.5cm}|}
 \hline
 \multicolumn{7}{|c|}{Polynomial chaos output} \\
 \hline
PC & p & q &r &$\epsilon_\text{LOO}$ & SFB  & SSB \\
 \hline
1   & 6    &0.85   &   4 &  $1.03\times 10^{-5}$  &  533  &  82\\
2   & 10    &0.60   &   3 &  $7.39\times 10^{-4}$  &  673  &  73\\
3   & 8   &0.60   &  3 &  $3.01\times 10^{-4}$   &  302  & 94\\
4   & 10    &0.65    &  4 & $5.2\times 10^{-3}$   &  960  & 48\\
5   & 10    &0.75    &  4 & $5.80\times 10^{-3}$   &  2304  & 88\\
6   & 6   &0.85    &  4 & $2.55\times 10^{-2}$   &  533 & 53\\
7   & 12    &0.55    &  3 & $4.04\times 10^{-2}$   &  771 & 59\\
8   & 10    &0.75    &  4 & $1.30\times 10^{-2}$   &  2304 & 60\\
9   & 14    &0.60    &  4 & $2.24\times 10^{-2}$   &  2150 & 53\\
 \hline
\end{tabular}
\label{PCEkev}
\end{table}

By retaining $99.999\%$ of the total variance of the dataset $\mathcal{D}$ when projecting it onto a lower-dimensional hyperplane during PCA, we identify $9$ principal components. Additional details and plots are available in Appendix \ref{appendix:PC}. Consequently, we create $9$ PCEs to emulate the weights these identified components. Each PCE undergoes an optimization process (basis adaptivity), with the parameters $p$ and $q$ being individually fine-tuned. The outcomes of this fine-tuning, including the LOO validation error ($\epsilon_{\text{LOO}}$) and the sizes of the full and sparse basis (denoted as SFB and SSB, respectively) are detailed in Table \ref{PCEkev}. In the absence of applying the hyperbolic truncation parameter $q$, the size of the full polynomial basis for any given degree $p$ is calculated according to Eq.\ \eqref{NumTerm}. For instance, considering the polynomial expansion for the first principal component with $p=6$ and given the input dimensionality of $M=7$, this calculation yields a basis size of $1716$ terms. Nevertheless, as clearly demonstrated in Table \ref{PCEkev}, the intervention of $q$ parameter has significantly reduced the basis size to $533$ terms. Subsequent use of LARS method for calculating the coefficients further reduced this number, yielding only $82$ of these terms having non-zero coefficients, with each of these polynomials depending on at most 4 variables, as indicated by $r = 4$. The graphical representation of non-zero coefficient for all the principal components  are illustrated in Appendix \ref{appendix:PC}.

\begin{figure}
    \includegraphics[width=0.45\textwidth]{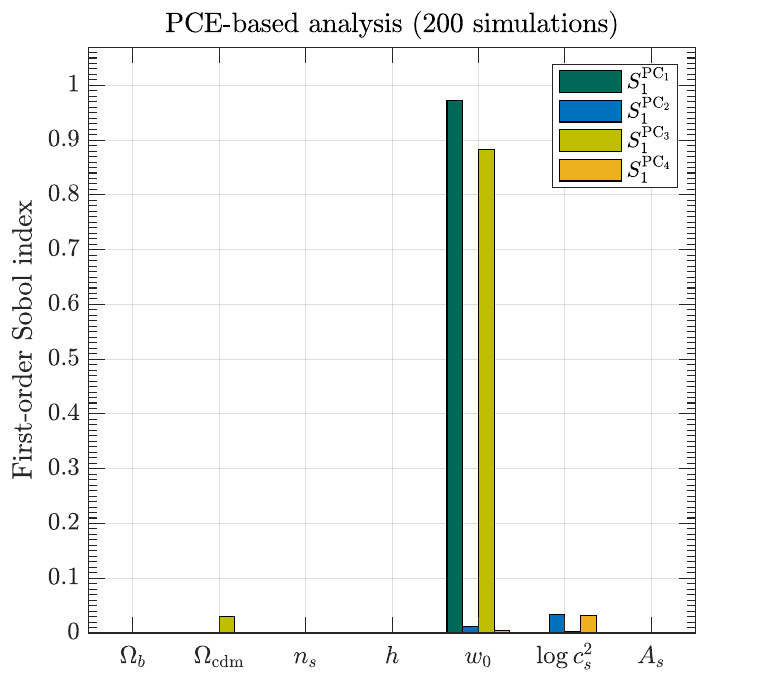}
    \caption{First-order Sobol indices for the first four principal components. The results are largely consistent with the mock emulator case, with a notable exception being the prominence of $c_s^2$ over $w_0$ in the second principal component, reflecting the specific impact of non-linear clustering of dark energy in \texttt{k-evolution} model.}
    \label{fig:kevSobol}
\end{figure}

We also conducted a first order Sobol index analysis for the first four principal components, with the findings illustrated in Fig.\ \ref{fig:kevSobol}. The outcomes closely match with those from the mock emulator scenario (Fig.\ \ref{fig:Sobol}), except for the distinct prominence of $c_s^2$ over $w_0$ in the second principal component. This difference can be attributed to the stronger influence of $c_s^2$ within \texttt{k-evolution} framework due to the non-linear clustering of dark energy.

\subsection{Emulator's performance}
\begin{figure}
    \centering
    \includegraphics[width=0.49\textwidth]{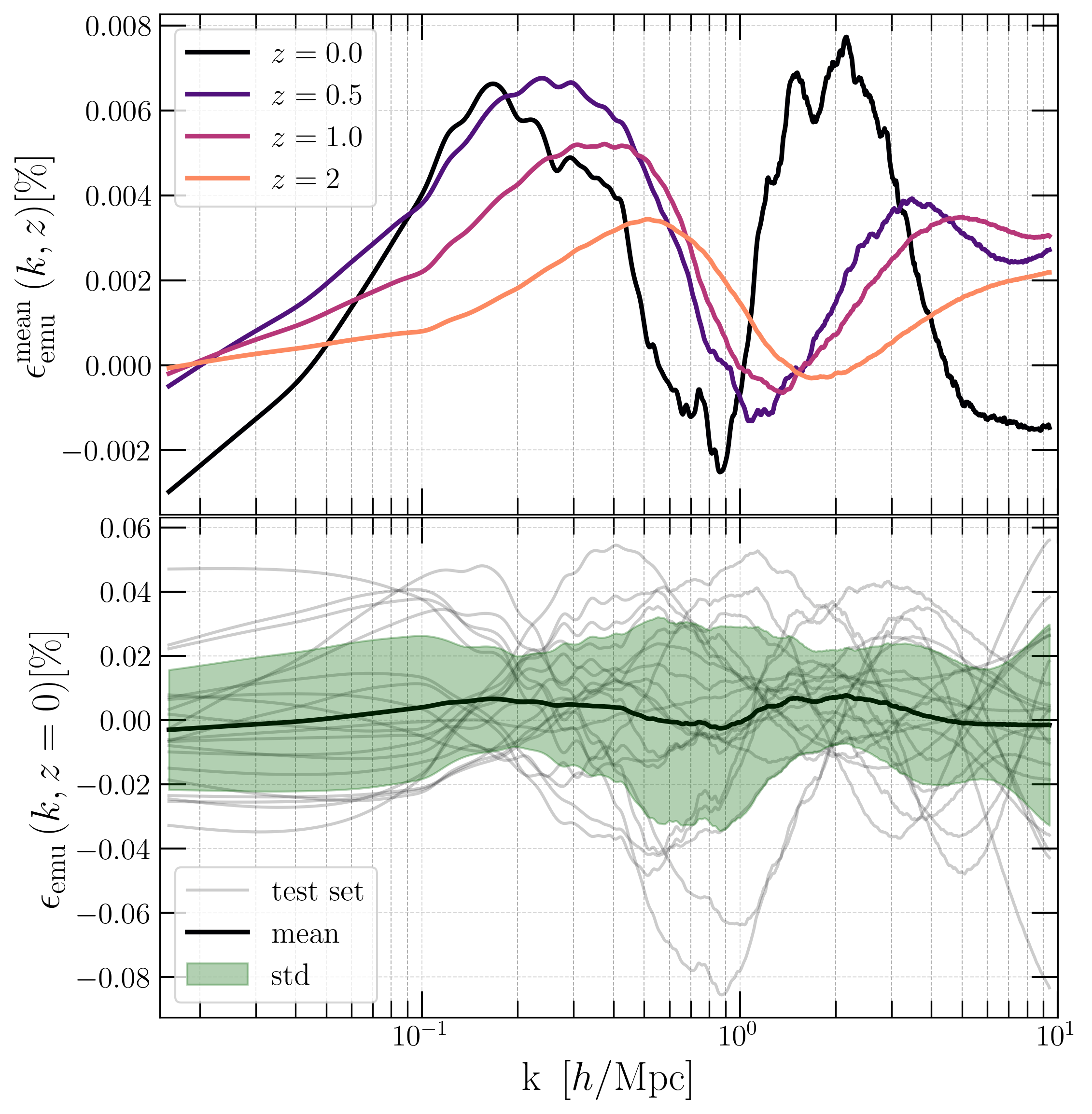}
    \caption{\textbf{Upper panel:} Average emulation error across $20$ cosmologies at different wavenumbers, observed at redshifts $z=0, 0.5, 1, 2$. \\
    \textbf{Lower panel:} Emulation error at $z=0$ for each set of 20 cosmologies (depicted in grey lines) together with the average emulation error at each $k$ (shown by the black line) and the standard deviation (represented in green). The maximum emulation error recorded is $\approx 0.08$ percent. }
    \label{fig:mean_emu}
\end{figure}

\begin{figure}
    \includegraphics[width=0.48\textwidth]{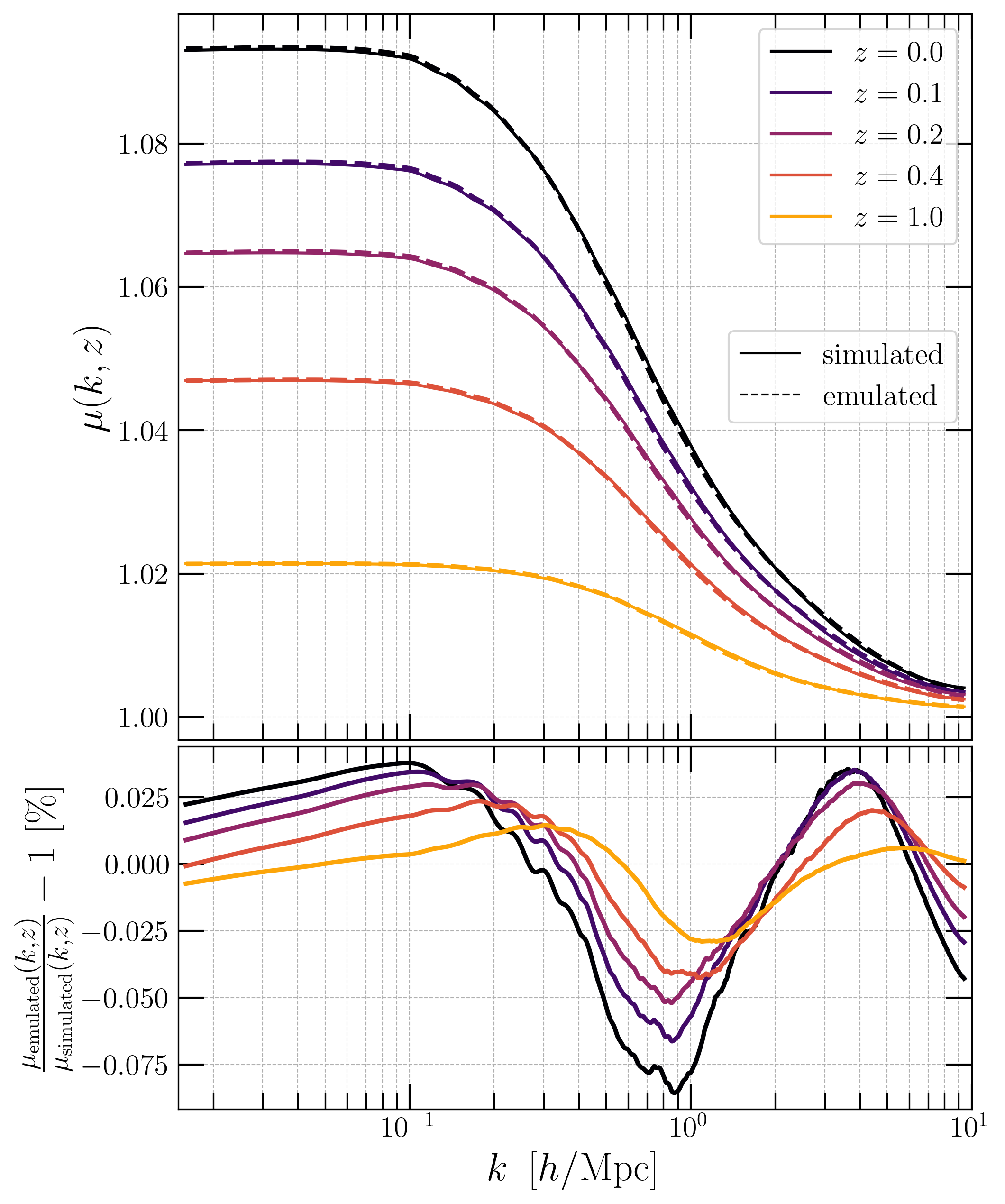}
    \caption{Comparison between simulated and emulated values of $\mu$ across various redshifts for the test set exhibiting the highest emulation error, approximately $0.08\%$ at $z=0$.}
    \label{fig:emulator_test}
\end{figure}

Due to computational constraints, we confined the evaluation of the emulator’s performance to $20$ test sets of cosmological parameters. These sets are randomly selected from a larger collection of $15\,000$ test sets previously generated during the configuration of the mock emulator. They are then subjected to \texttt{k-evolution} simulation, employing the same number of particles and volume size as those used in the training process. In Fig.\ \ref{fig:mean_emu}, we show the average emulation error over the $20$ test sets at each wavenumber for different redshifts (upper panel), alongside the emulation errors for all $20$ test set cosmologies at $z=0$ (lower panel). From this group, we identified the cosmology corresponding to the highest emulation error (approximately $0.08\%$) and compare the simulated and emulated $\mu$ at different redshifts. The comparative results are depicted in Fig.\ \ref{fig:emulator_test}.

In terms of speed, the emulator—implemented within a Python wrapper—takes approximately $0.2$ seconds ($5.36 \times 10^{-5}$ core hours) per evaluation on a regular laptop, compared to the $2.56 \times 10^{3}$ core hours required for a simulation at the same resolution. This corresponds to a speed-up factor of nearly fifty million.

\section{Reconstructing the potential power spectrum}
\label{sec:reproduce}
\begin{figure*}
    \includegraphics[width=0.9\textwidth]{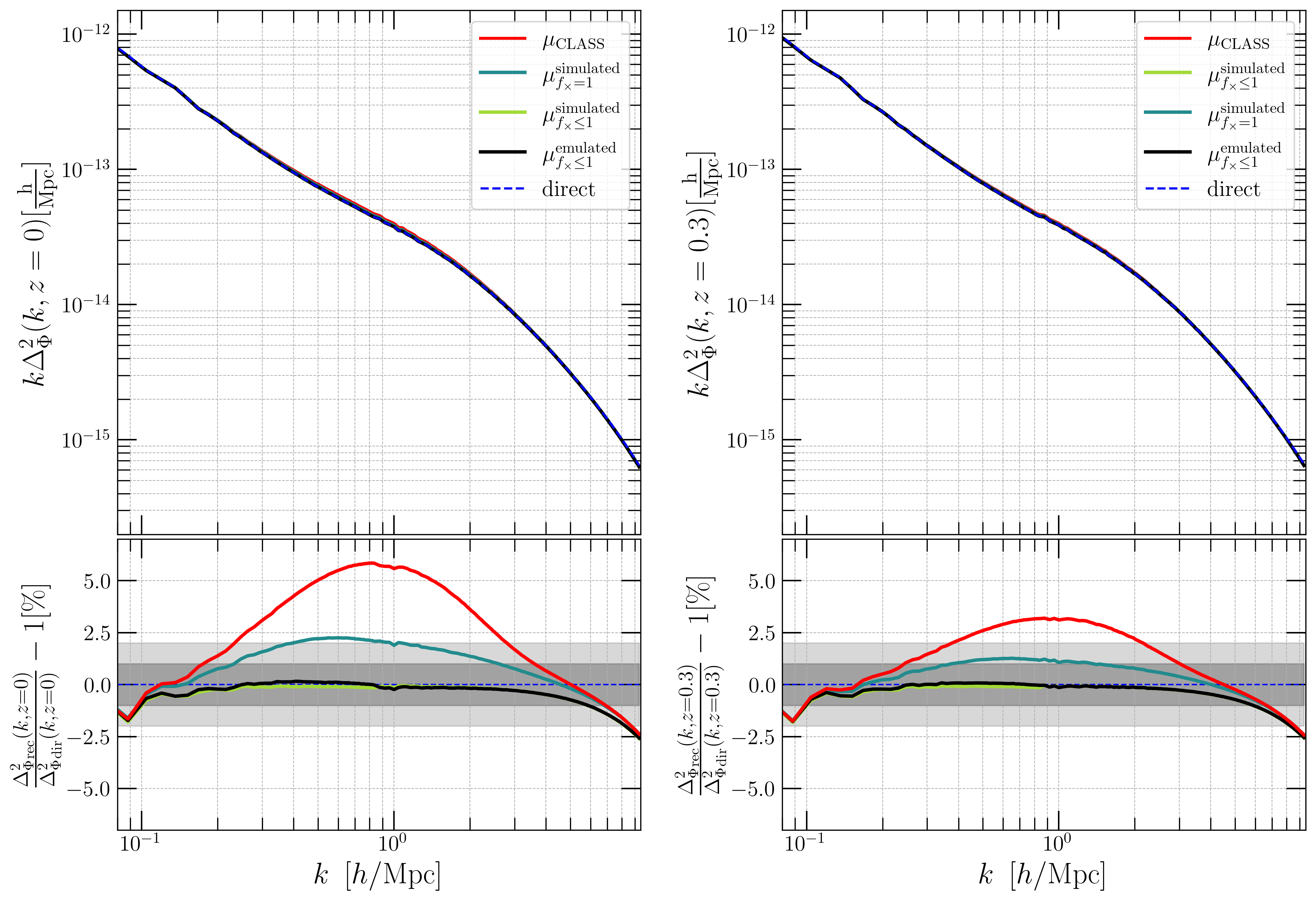}
    \caption{Reconstruction of the potential power spectrum using Eq.\ \eqref{reproduce} through the use of the $\mu$ function and the matter power spectrum at $z=0$ (left panel) and $z=0.3$ (right panel). Employing both the simulated and emulated $\mu$ function in which the suppression in cross-correlation between dark matter and dark energy at small scales is considered (respectively denoted as $\mu^\text{simulated}_{f_\times \leq 1}$ and $\mu^\text{emulated}_{f_\times \leq 1}$), results in consistent outcomes compared to the direct output of the potential power spectrum. This consistency is observed within the wavenumber range unaffected by cosmic variance and resolution effects at both $z=0$ and $z=0.3$. At scales affected by these factor, the results surpass the $1\%$ threshold. Conversely, when using $\mu^\text{simulated}_{f_\times = 1}$, i.e. assuming a full correlation between dark matter and dark energy, the deviation from the direct output reaches a maximum of $\sim2.5\%$ at $z=0$, while exhibiting almost $1\%$ maximum deviation at $z=0.3$ due to the less effective clustering of dark energy.}
    \label{fig:potentials_rec}
\end{figure*}
In this section, we explain the process of determining the potential power spectrum by using the $\mu$ function together with the matter power spectrum, showing how accurate estimations of both can enable a precise calculation of the potential power spectrum.

To derive the potential power spectrum, one may rearrange Eq.\ \eqref{mu-0} such that
\begin{equation}
    k\Delta^2_\Phi =  \frac{\mu(k,z)^2 (4\pi G_N a^2\bar\rho_m)^2 \Delta^2_m}{k^3} ,
    \label{reproduce}
\end{equation}
where $\Delta^2_\Phi$ and $\Delta^2_m$ are respectively dimensionless potential and matter power spectrum given by
\begin{equation}
    \Delta_X^2(k) =  \frac{k^3}{2\pi^2}P_X(k).
\end{equation}
As demonstrated in Fig.\ \ref{fig:potentials_rec}, efforts to recreate the true potential power spectrum based on the cosmological parameters of Table \ref{table:cosmoparams}, reveal that using the \texttt{CLASS}-based $\mu$ function ($\mu_\text{CLASS}$) results in a maximum information loss of more than $5\%$ at $z=0$ compared to a direct calculation of the true potential power spectrum obtained with \texttt{k-evolution} code. Even applying a simulated $\mu$ function that disregards the true cross-correlation between dark energy and dark matter, $\mu^\text{simulated}_{f_\times = 1}$, a maximum deviation of approximately $2\%$ at the same redshift is observed. In contrast, consistent reproduction of $\Phi$ power spectrum is only achieved when employing the $\mu$ function that accounts for the true cross-correlation, i.e. $\mu^\text{simulated}_{f_\times \leq 1}$. At $z=0.3$, both $\mu_\text{CLASS}$ and $\mu^\text{simulated}_{f_\times = 1}$ demonstrate less deviations compared to the scenario at $z=0$. This is because, at higher redshifts, the less pronounced clustering of dark energy results in  a stronger correlation with dark matter. \\
All the results shown in Fig.\ \ref{fig:potentials_rec}, including the computation of $\mu$ function and the direct computation of the potential power spectrum, are derived from the combination of the two mentioned Flagship simulations in Section \ref{sec:ConvTest}.  It is evident that reproducing the potential power spectrum, using the high-resolution $\mu^\text{simulated}_{f_\times \leq 1}$ in one instance and the emulated $\mu^\text{emulated}_{f_\times \leq 1}$ in another, has produced nearly identical results. This supports our choice to opt for the simulation settings $N_\text{grid} = 1200^3$ and $L = 400 ~h^{-1}~ \text{Mpc}$ in the convergence test investigation and indicates that an exact predicted value of $\mu$ function (using our emulator) and the matter power spectrum (using, for example, \texttt{EuclidEmulator}) will result in a precise reproduction of the potential power spectrum through the Eq.\ \eqref{reproduce}.

The reproduction error using the emulated $\mu$ function remains well below the $1\%$ limit across the
wavenumber range unaffected by cosmic variance and resolution effects, i.e. from $k = 0.02 h \, \text{Mpc}^{-1}$ to $k = 6 h \, \text{Mpc}^{-1}$. However, it exceeds the $1\%$ limit on small scales due to resolution constraints and on large scales due to cosmic variance. These deviations might be attributed to how differently cosmic variance and resolution effects are incorporated into the potential and dark energy spectra by different solvers.
To further investigate this issue, we conducted simulations at varying resolutions
and calculated the $\mu$ function using Eq.\ \eqref{mu-0}. We then compared these results within their reliable range --not skewed by cosmic variance and resolution effects-- to the $\mu$ output from a single fixed-resolution simulation obtained from Eq.\ \eqref{eq:mu-corr}. By observing that the results from the variable-resolution simulations using Eq.\ \eqref{eq:mu-corr} match with the one obtained from Eq.\ \eqref{eq:mu-corr}, we then concluded that using both the matter power spectrum and $\mu$ function obtained from Eq.\ \eqref{mu-1}, gives a more accurate estimation of potential power spectrum compared to the direct output provided by the simulation.

\section{Conclusions}

We develop an emulator to accurately model the non-linear clustering effects of the $k$-essence dark energy, using the PCE approach implemented in the \texttt{UQLab} software. The emulator is built on a seven dimensional parameter space and is based on \texttt{k-evolution} simulations using $200$ training samples. We use a mock (\texttt{Halofit}-based) emulator to study the optimal emulator configuration, and find
a maximum emulation error of $0.15\%$ using only $11$ principal components. This was assessed against the outcome of $15\,000$ sets of cosmological parameters that were randomly sampled within a hypersphere of radius 1, inscribed in the normalized parameter space. By applying the insights gained from the construction of the mock emulator, we find a $0.08\%$ maximum emulation error in the case of the \texttt{k-evolution}-based emulator (projected onto $9$ principal components) which is assessed against only $20$ test sets due to the limited computational budget. The emulator is executed in a wall time of $\sim 0.2 s$ on a usual laptop which is comparable to the case of \texttt{EuclidEmulator1} ($\sim 0.4 s$ per evaluation) and \texttt{EuclidEmulator2} ($\sim 0.3 s$ per evaluation) in which the same emulation method is used.

Using the polynomial expansions obtained for the emulator, we perform a sensitivity analysis based on Sobol indices, which in turn reveals the dominant influence of $w_0$ and $c_s^2$ on $\mu(k,z)$. Through the same analysis, we also observed that the sum of neutrino masses have the least impact on the $\mu$ function among other cosmological parameters.

Our emulator predicts the $\mu$ function that encodes the amount of clustering in dark energy relative to the dark matter. Since it is based on the ratio of two power spectra, the $\mu$ function suffers much less from cosmic variance and resolution effects than the power spectra themselves. This is a crucial advantage, since it significantly reduces the requirements for particle numbers and simulation volume for a given accuracy.

One important application of our emulator is the accurate reconstruction of the lensing potential power spectrum from a matter power spectrum. To obtain the correct power spectrum on non-linear scales, one needs
to account for the true cross-correlation between dark energy and dark matter in the definition of $\mu$. Overlooking this aspect leads to a non-negligible deviation from the true potential power spectrum at low redshifts. This is particularly critical for the lensing signal, where errors accumulate due to its integrating effect. For this reason, we built two emulators, one that includes the true cross-correlations between dark energy and dark matter and one that assumes a full correlation between the two at all scales and redshifts. We demonstrate explicitly that the former, i.e. $\mu^\text{emulated}_{f_\times \leq 1}$ , allows to recover the $\Phi$ power spectrum to high accuracy from the matter power spectrum.

The emulator is available for public use at \url{https://github.com/anourizo/k-emulator}. The repository includes the data for polynomial bases and their respective coefficients for both variants of the $\mu$ function, with $f_\times = 1$ and $f_\times \leq 1$. Additionally, it includes the Python wrapper and a DEMO Jupyter notebook that guide users through the process of using the emulator to generate emulated $\mu$ function.

\section*{Acknowledgements}

ARN and MK acknowledge funding from the Swiss National Science Foundation.
FH acknowledges the Research Council of Norway and the resources provided by 
UNINETT Sigma2 -- the National Infrastructure for High Performance Computing and 
Data Storage in Norway. FH would also like to thank Elif Ilhan, \'Alvaro de la Cruz-Dombriz, Peter Dunsby, Roy Maartens and James Adam for their support and hospitality during the research visit to Cape Town. 
ARN would like to thank the Vahabzadeh Foundation for their continuous financial support. This work is also supported by a grant from the Swiss National Supercomputing Centre (CSCS) under project ID s1051.
Most of the computations were performed at the University of Geneva using Baobab and Yggdrasil HPC services.

\section*{Data Availability}
The emulator code and data is publicly available at \url{https://github.com/anourizo/k-emulator} .
 



\bibliographystyle{mnras}
\bibliography{bibliography} 




\appendix

\section{Analysis of variance (ANOVA) based on polynomial chaos expansion}

\subsection{Hoeffding-Sobol decomposition}
We define $g({\bm x})$ as a square-integrable function representing a computational model $Y$, where ${\bm x} = \{x_1, x_2, \ldots, x_M\}$ is an $M$-dimensional vector of independent variables within the support $\mathcal{D}_{\bm x}$ and probability density function $f$, given by
\begin{equation}
\mathcal{D}_{\bm x} := \{ 0 \leq x_i \leq 1 | i \in 1, \ldots, M; M \in \mathbb{N} \}, \quad f({\bm x}) = 1.
\end{equation}

The goal is to analyze how sensitive $g({\bm x})$ is to each input parameter, their pairs, triplets, and so on.
According to \cite{Sobol2001} , the function $g(\bm{x})$ can be decomposed into components of increasing dimensions, expressed as follow:
\begin{align}
g({\bm x}) = g_{0} &+ \sum_{i=1}^{M} g_{i}\left(x_{i}\right) 
+\sum_{1 \leq i<j \leq M} g_{i j}\left(x_{i}, x_{j}\right) \nonumber  \\ 
&+\sum_{1 \leq i_1< \ldots < i_s \leq M} g_{i_1 \ldots i_s}\left(x_{i_1}, \ldots, x_{i_s}\right) 
 +\ldots + g_{1, \ldots, M}\left({\bm x}\right) .
\label{Hoeff}
\end{align}

In the above expression $g_0$ is a constant term and is equal to the expected value of $g(\bm x)$. It is also assumed that the integrals of the summands $g_{i_1 \ldots i_s}$ with respect to any of their own variables vanish
 \begin{equation} 
\int_{0}^{1} g_{ i_{1}, \ldots, i_{s}}\left(x_{i_{1}}, \ldots, x_{i_{s}}\right) d x_{i_{k}}=0 
, ~1 \leq k \leq s , ~1 \leq i_1\leq ...i_s\leq M ,
\label{own}
\end{equation} 
which also yields to the fact that the summands in the expression \eqref{Hoeff} are mutually orthogonal in the integration space, and can be calculated in a recursive approach as proposed in \cite{sensitivity2022}:
\begin{equation}
 \begin{aligned}
g_i\left(x_i\right) &=\int_0^1 \cdots \int_0^1 g(\boldsymbol{x}) d \boldsymbol{x}_{\sim i}-g_0  =
\mathbb{E}(Y|x_i)-g_0 ,\\  
g_{i j}\left(x_i, x_j\right) &=\int_0^1 \cdots \int_0^1 g(\boldsymbol{x}) 
d \boldsymbol{x}_{\sim(i j)}-g_0-g_i\left(x_i\right)-g_j\left(x_j\right)\\
&=  \mathbb{E}(Y|x_i,x_j)-g_0 -g_i - g_j .\\ \nonumber
\label{recursive}
\end{aligned}
\end{equation} 
where $\sim$ stands for the excluded variables, i.e.
\begin{equation}
     d \boldsymbol{x}_{\sim i} = (dx_1,dx_2,...,dx_{i-1},dx_{i+1},...,dx_{M-1},dx_M) .
\end{equation}
The orthogonality of these components, as indicated by this approach, also leads to the {\textit{unique}} nature of the decomposition \eqref{Hoeff}.

\subsection{Sobol indices}
Building upon the preceding results, the total variance of $Y$ is expressed as
\begin{equation}
\text{D}=\text{Var}{[Y]}=\int_{\mathcal{D}_{\boldsymbol{x}}} g^{2}(\boldsymbol{x}) d \boldsymbol{x}-g_{0}^{2} .
\label{totvar}
\end{equation}
Integrating the square of the expression (\ref{Hoeff}) over $\mathcal{D}_{\bm x}$ allows the decomposition of the total variance $\text{D}$ into partial variances
\begin{equation}
\text{D}=\sum_{i=1}^{M} \text{D}_{i}+\sum_{1 \leqslant i<j \leqslant M} \text{D}_{i j}+\cdots+
\text{D}_{1,2, \ldots, M} ,
\label{parvar}
\end{equation} 
where the partial variances are defined as
 \begin{equation}
\text{D}_{i_{1}, \ldots, i_{s}}=\int_{0}^{1} \ldots \int_{0}^{1} g_{i_{1}, \ldots, i_{s}}^{2}
\left(x_{i_{1}}, \ldots, x_{i_{s}}\right) d x_{i_{1}} \ldots d x_{i_{s}} .
\end{equation}
The Sobol indices are then introduced as
\begin{equation} 
S_{i_{1}, \ldots, i_{s}}=\frac{\text{D}_{i_{1}, \ldots, i_{s}}}{\text{D}} ,
\label{sobol_ind}
\end{equation}
and according to the equation (\ref{parvar}), they follow the constraint
\begin{equation}
\sum_{i=1}^{M} S_{i} + \sum_{1 \leqslant i<j \leqslant M} S_{i j} + \cdots + S_{1,2, \ldots, M} = 1 .
\label{indices}
\end{equation}

The first-order Sobol indices $S_i$ (main effects) characterize the fractional contribution of each individual input variable to the total variance, while the second-order Sobol indices $S_{ij}$, where $i \neq j$, indicate the fractional contribution of each pair of input variables to the overall output variance. The interpretation extends similarly to $M$-th order Sobol indices.

\subsection{PCE-based Sobol indices}
Typically, calculating Sobol indices involves conducting Monte Carlo simulations. Nonetheless, the Eq.\ \eqref{Hoeff} contains a total number of terms represented by
\begin{equation}
\sum_{l = 1}^{M}  
\begin{pmatrix} 
M \\
l 
\end{pmatrix} = 2^M-1 .
\end{equation}
This implies that a minimum of $2^M$ Monte Carlo integrals is necessary for estimating these indices. However, this approach becomes impractical for models with a large number of dimensions ($M$). To resolve this issue, we can turn to PCE for calculating Sobol indices, an approach suggested by \cite{SUDRET2008964}.

Based on the framework proposed in \cite{sensitivity2022}, we can develop the concept of PCE-based sensitivity analysis as follows: introducing the set $\mathcal{U} = \{1,...,M\}$ and its subset $\bm u \subset \mathcal{U}$, the Hoeffding-Sobol decomposition \eqref{Hoeff} can be rewritten as follow 
\begin{equation} 
 g({\bm x}) = g_0 + \sum_{\bm u \subset \mathcal{U}}g_{\bm u}({\bm x}_{\bm u}) .
 \label{reHoeff}
\end{equation}
In this equation, $g_{\bm u}$ denotes the partial effect of the parameters within the subvector $\bm x_{\bm u} = (x_i)_{i \in \bm u}$. 

On the other hand, the truncated polynomial expansion, Eq.\ \eqref{finpce}, can be rearranged to form an expansion in terms of summands of increasing order. This is done by imposing a new condition on
the finite index set $\mathcal{A}^{M,p,r}_q$, such that for a non-empty
set $\bm v \subset \mathcal{U}$, 
$\mathcal{A}_{\bm v}$ is defined as:
\begin{equation} 
\mathcal{A}_{\bm v}:=\left\{{\bm\alpha}\in \mathcal{A}^{M,p,r}_q, 
 \begin{pmatrix}
\alpha_k>0, \forall k=1, \ldots, M \mid k \in{\bm v} \\
\alpha_k=0, \forall k=1, \ldots, M \mid k \notin{\bm v}
\end{pmatrix}
\right\} .
\end{equation}
Subsequently, the rearranged truncated PCE is represented as,
\begin{equation}
 Y = \beta_0 + \sum\limits_{\bm v \subset \mathcal{U}}
 \sum\limits_{\bm{\alpha} \in \mathcal{A}_{\bm v}} \beta_\alpha \Psi_{\alpha}(\boldsymbol{x}) .
 \label{reordered}
\end{equation} 
Since the Hoeffding-Sobol decomposition is unique, comparing Eqs.\ \eqref{reHoeff} and \eqref{reordered} results in
\begin{equation}
g_{\bm{u}}(\bm{x}_{\bm{u}}) = 
 \sum\limits_{\bm{\alpha} \in \mathcal{A}_{\bm v}} \beta_\alpha \Psi_{\alpha}(\boldsymbol{x}) .
\end{equation}
Given the orthonormality of multivariate polynomial basis, both the partial and total variances can be attributed to the squares of the PCE coefficients
\begin{equation}
\begin{aligned}
 \text{Var}[Y] &= \sum_{\bm \alpha \in \mathcal{A}^{M,p,r}_q} \beta_{\bm \alpha}^2  \\
 \text{Var}[g_{\bm{u}}(\bm{x}_{\bm{u}})] &= 
 \sum_{\bm \alpha \in \mathcal{A}_{\bm v}} \beta_{\bm \alpha}^2 , \\ 
 \end{aligned}
\end{equation}
Consequently, the corresponding Sobol indices can be represented as follow:
\begin{equation}
 S_{\bm v} =\frac{ \text{Var}[g_{\bm{u}}(\bm{x}_{\bm{u}})]}{\text{Var}[Y]}  = 
 \frac{\sum\limits_{\bm \alpha \in \mathcal{A}_{\bm v}} \beta_{\bm \alpha}^2}
 {\sum\limits_{\bm \alpha \in \mathcal{A}^{M,p,r}_q} \beta_{\bm \alpha}^2} .
\end{equation}

\section{Relative emulation error in 2D planes}

\begin{figure*}
    \centering
    \includegraphics[width=0.95\textwidth]{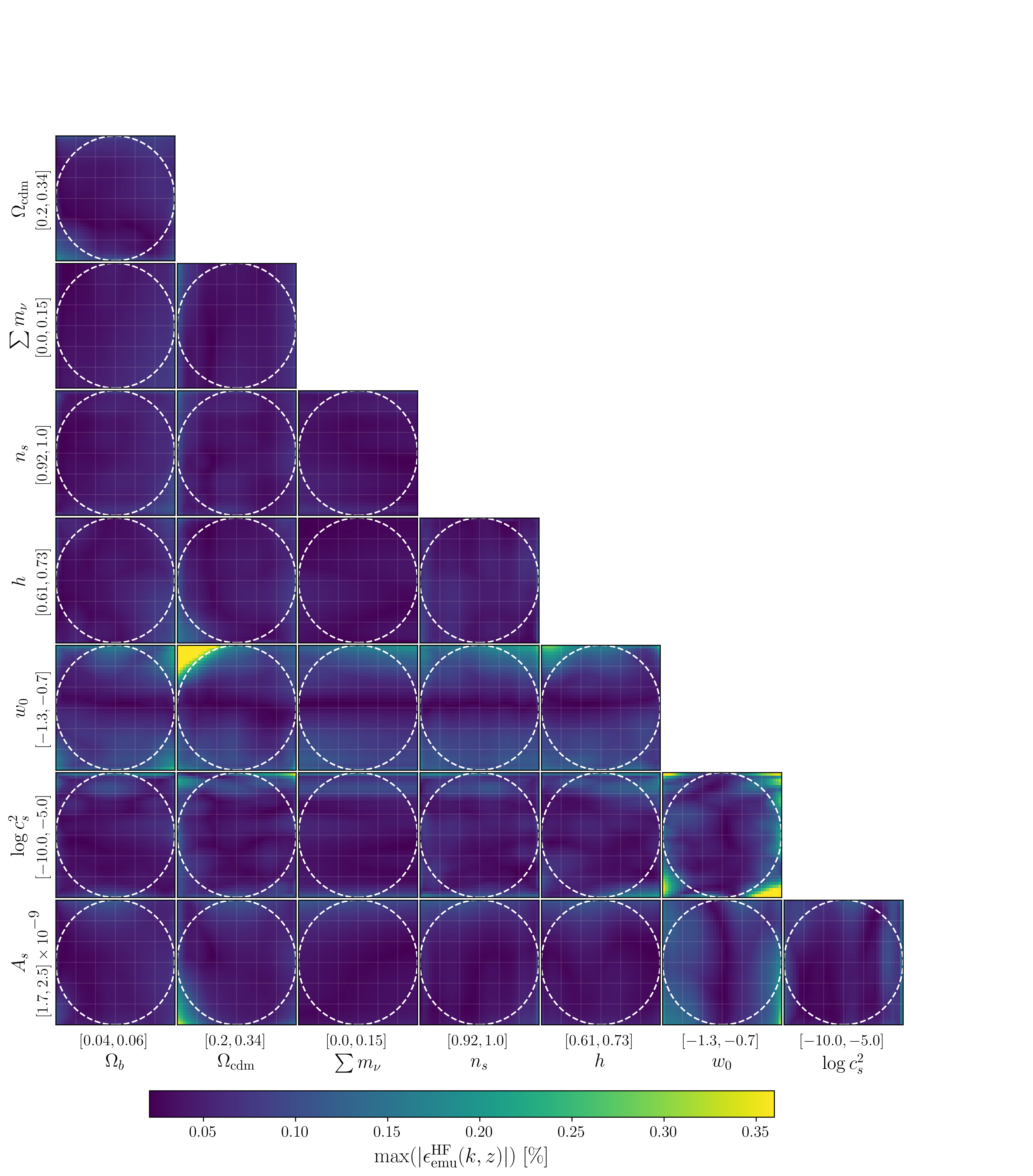}
    \caption{Heat maps displaying the emulator's performance across 28 distinct 2D planes, each containing $1600$ grid test samples.}
    \label{fig:planes}
\end{figure*}

In Fig.\ \ref{fig:planes} we show the maximum emulation error of the mock emulator across $28$ different 2D planes, with each plane containing $1\,600$ test samples. We see that the most significant deviations appear for extreme choices (corners) of the dark-energy related parameters $w_0$ and $c_s^2$ as well as $\Omega_{\rm cdm}$ which indirectly controls $\Omega_{\rm DE}$. It should be mentioned that, even though these are the most significant deviations, they still correspond to low emulation errors.

\section{Principal components of the response data}
\label{appendix:PC}
\begin{figure*}
\hspace{-0.2cm}
    \includegraphics[width=0.9\textwidth]{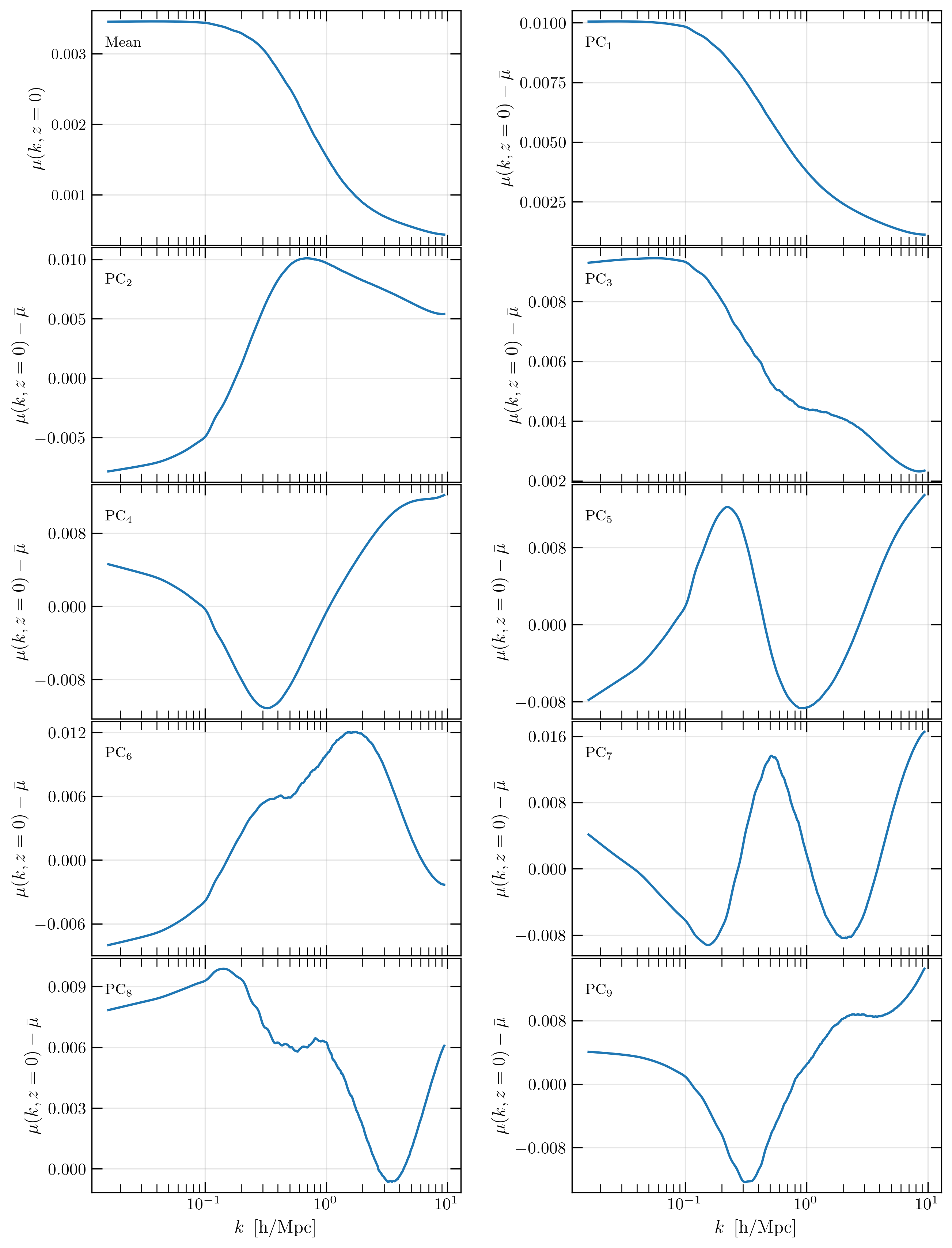}
    \caption{Illustration of the mean of response data and eigenvectors obtained from the covariance matrix of the response data. These principal components, after being scaled by their corresponding eigenvalues generated through emulation, are aggregated with the mean. This combination forms the emulated $\mu$ function. }
    \label{fig:PCBasis}
\end{figure*}

\begin{figure*}
\centering
\begin{subfigure}{0.33\textwidth}
  \centering
  \includegraphics[width=\linewidth]{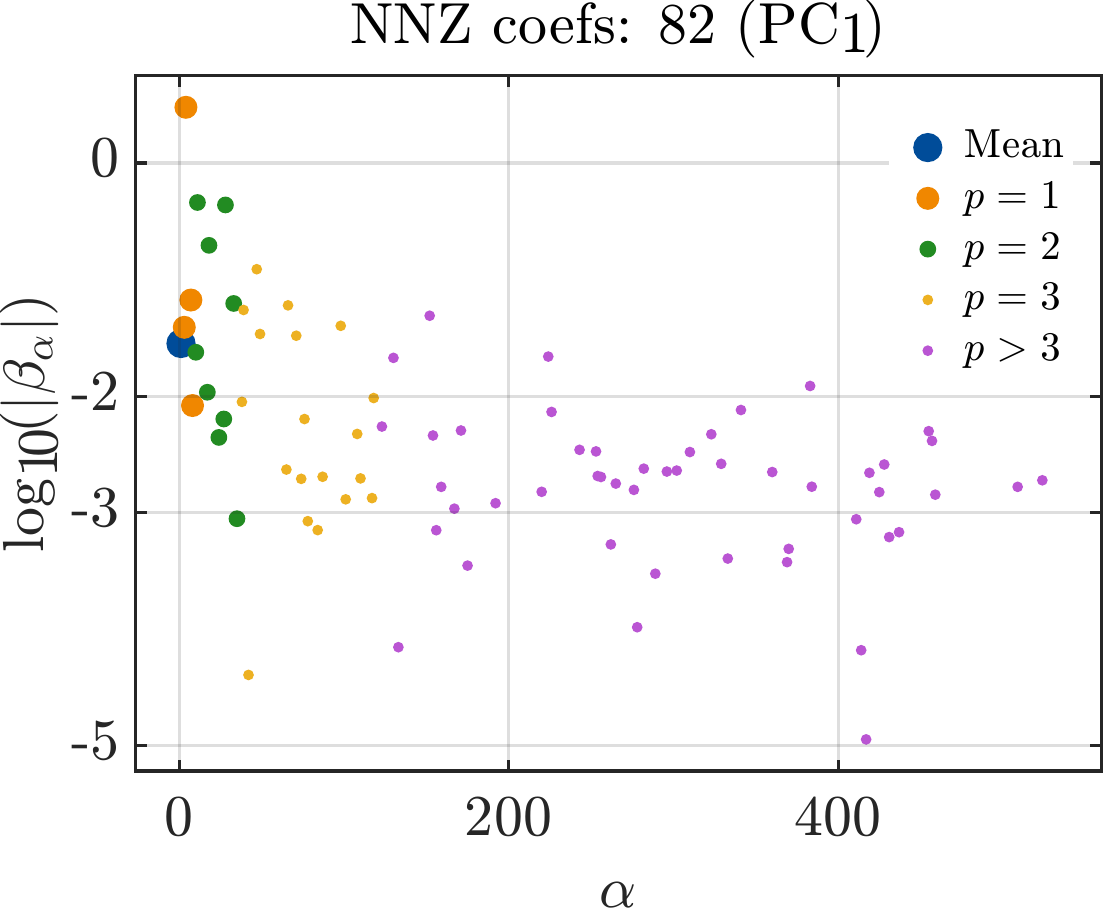}
  \vspace{5pt}
\end{subfigure}%
\hfill
\begin{subfigure}{0.33\textwidth}
  \centering
  \includegraphics[width=\linewidth]{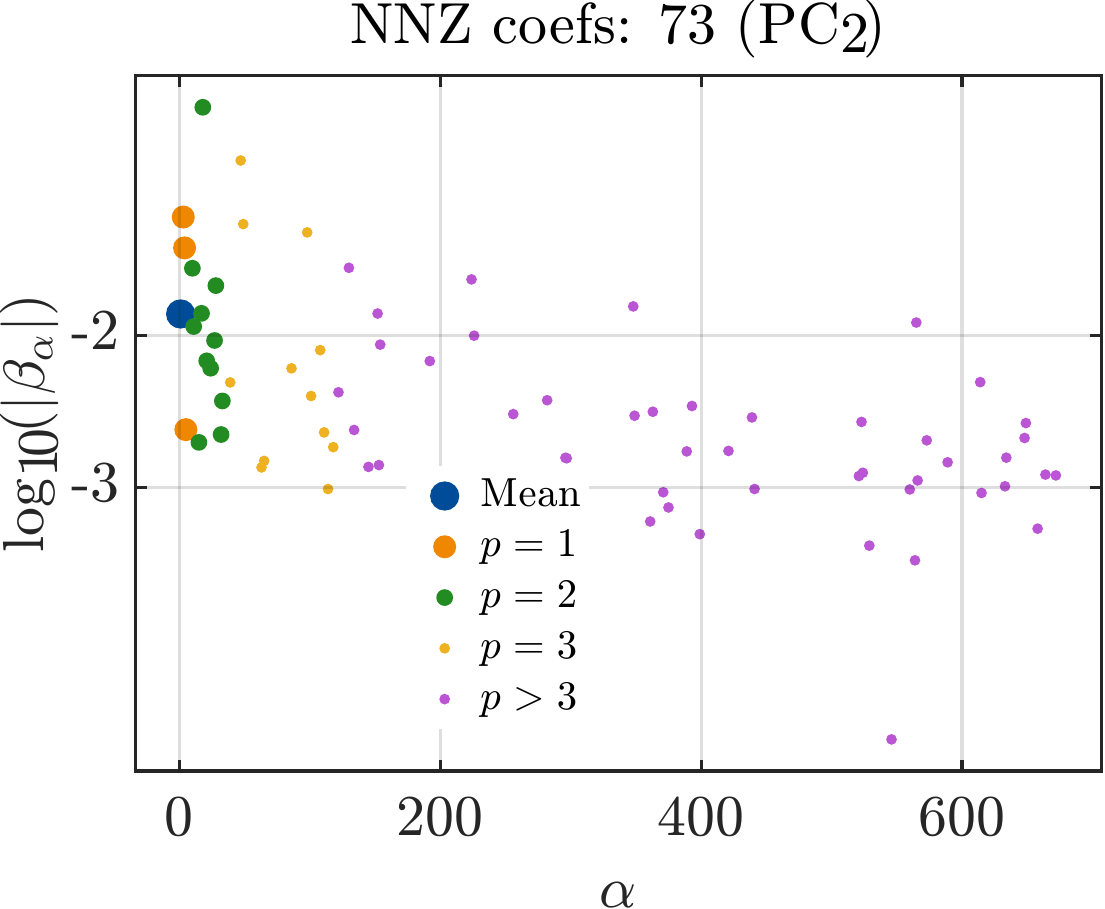}
  \vspace{5pt}
\end{subfigure}%
\hfill
\begin{subfigure}{0.33\textwidth}
  \centering
  \includegraphics[width=\linewidth]{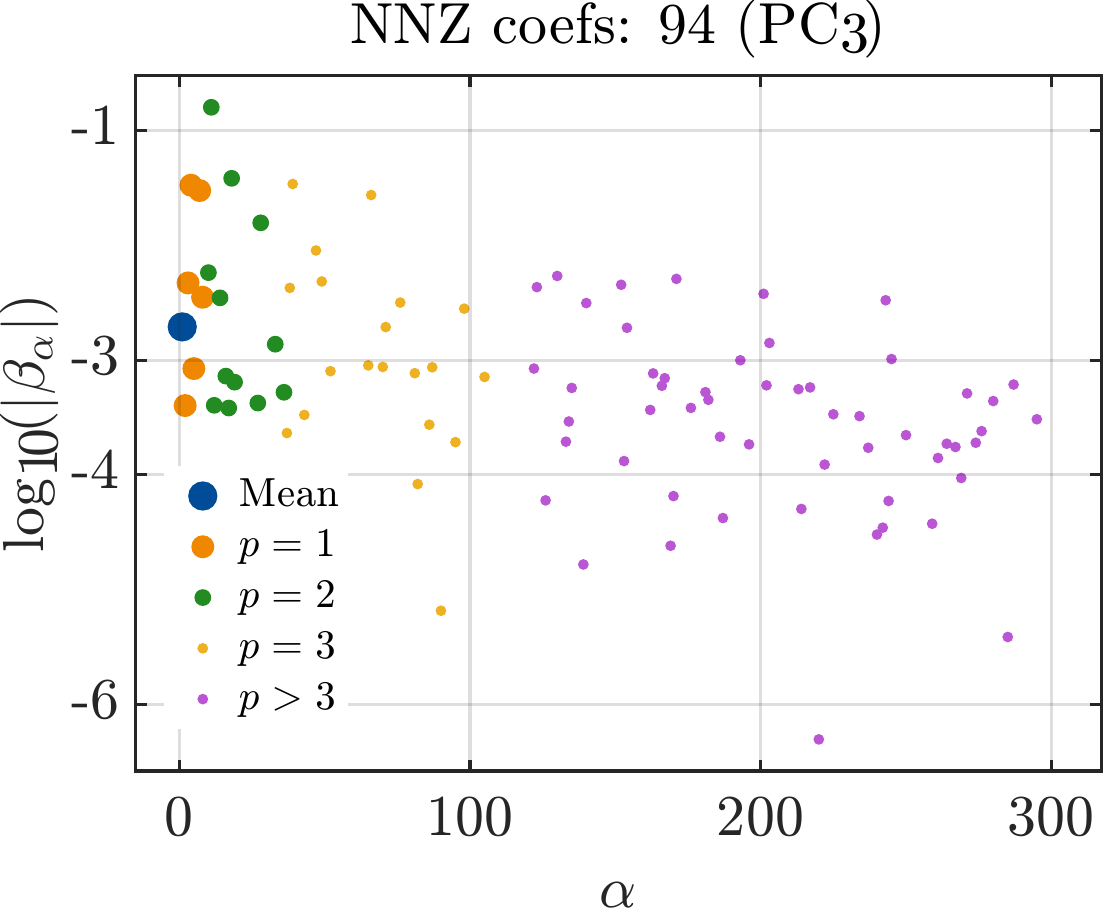}
  \vspace{5pt}
\end{subfigure}

\begin{subfigure}{0.33\textwidth}
  \centering
  \includegraphics[width=\linewidth]{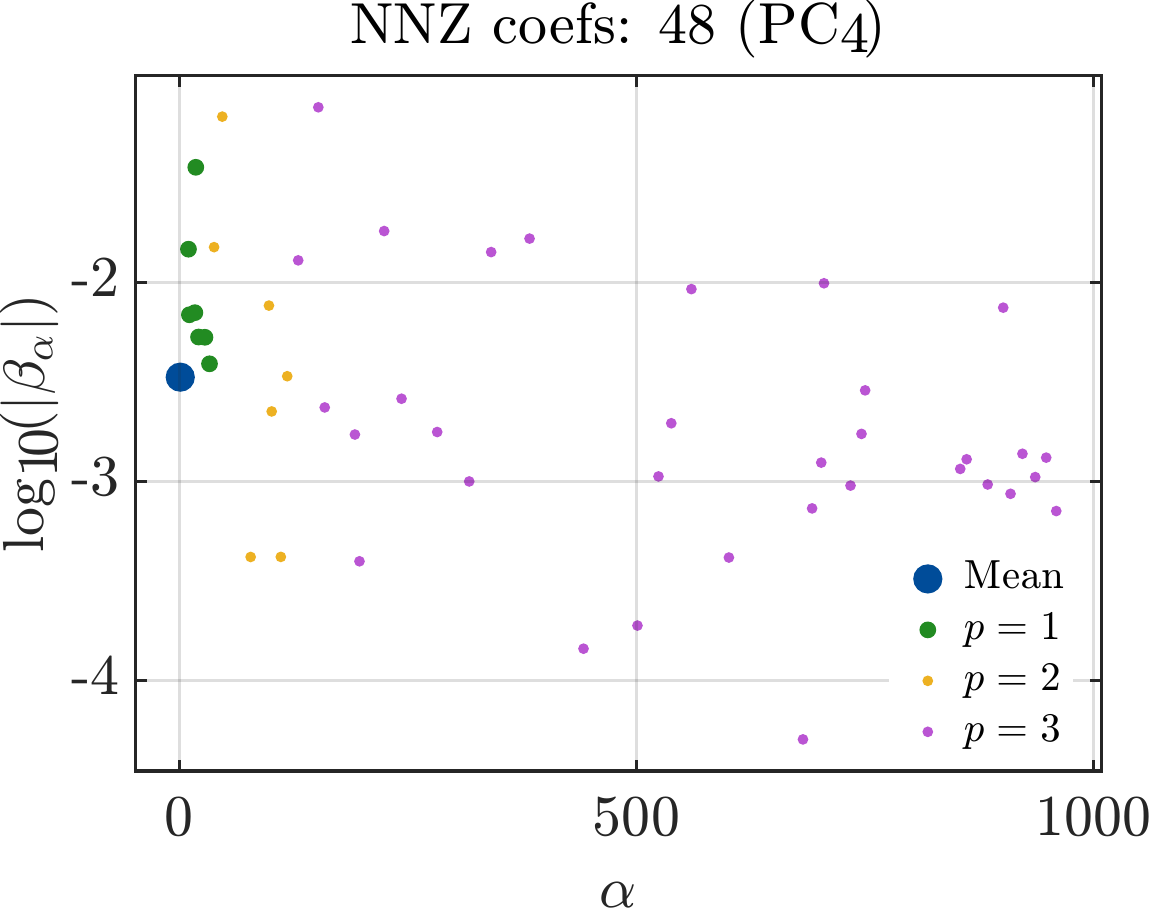}
  \vspace{5pt}
\end{subfigure}%
\hfill
\begin{subfigure}{0.33\textwidth}
  \centering
  \includegraphics[width=\linewidth]{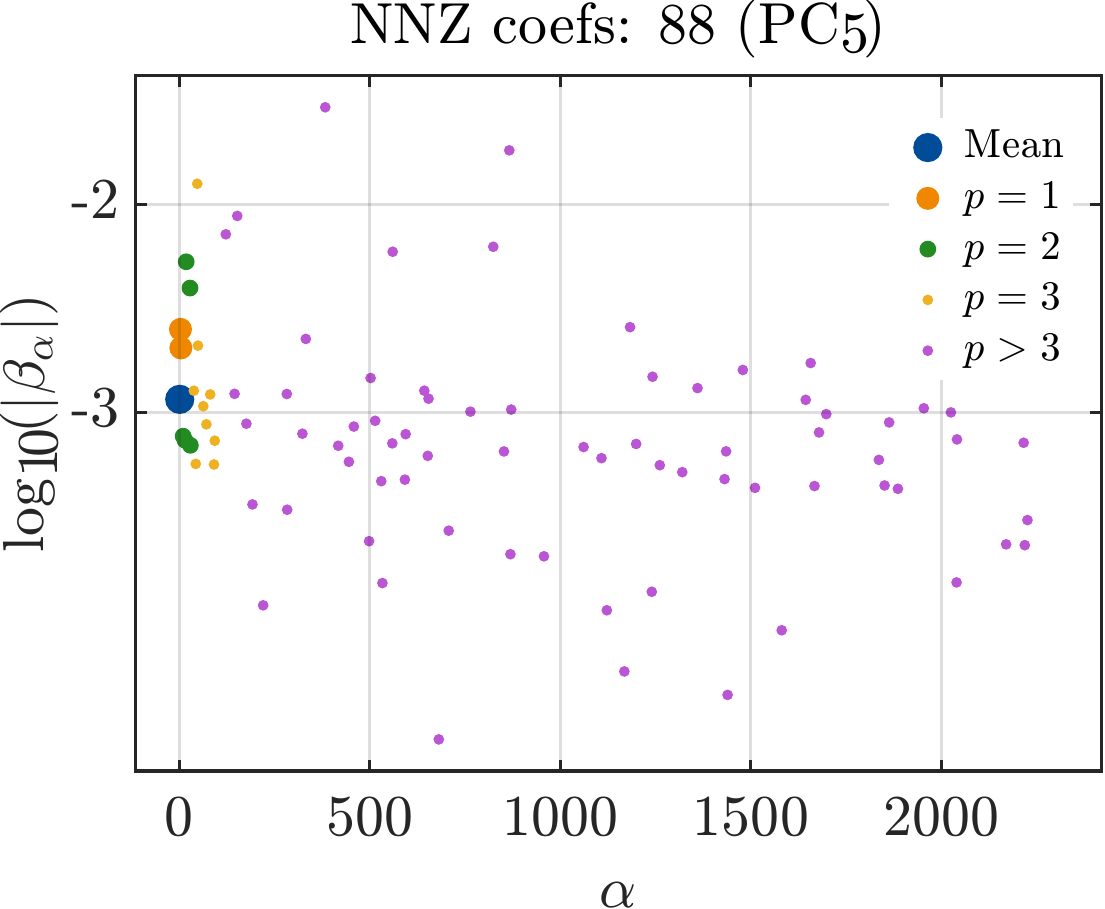}
  \vspace{5pt}
\end{subfigure}%
\hfill
\begin{subfigure}{0.33\textwidth}
  \centering
  \includegraphics[width=\linewidth]{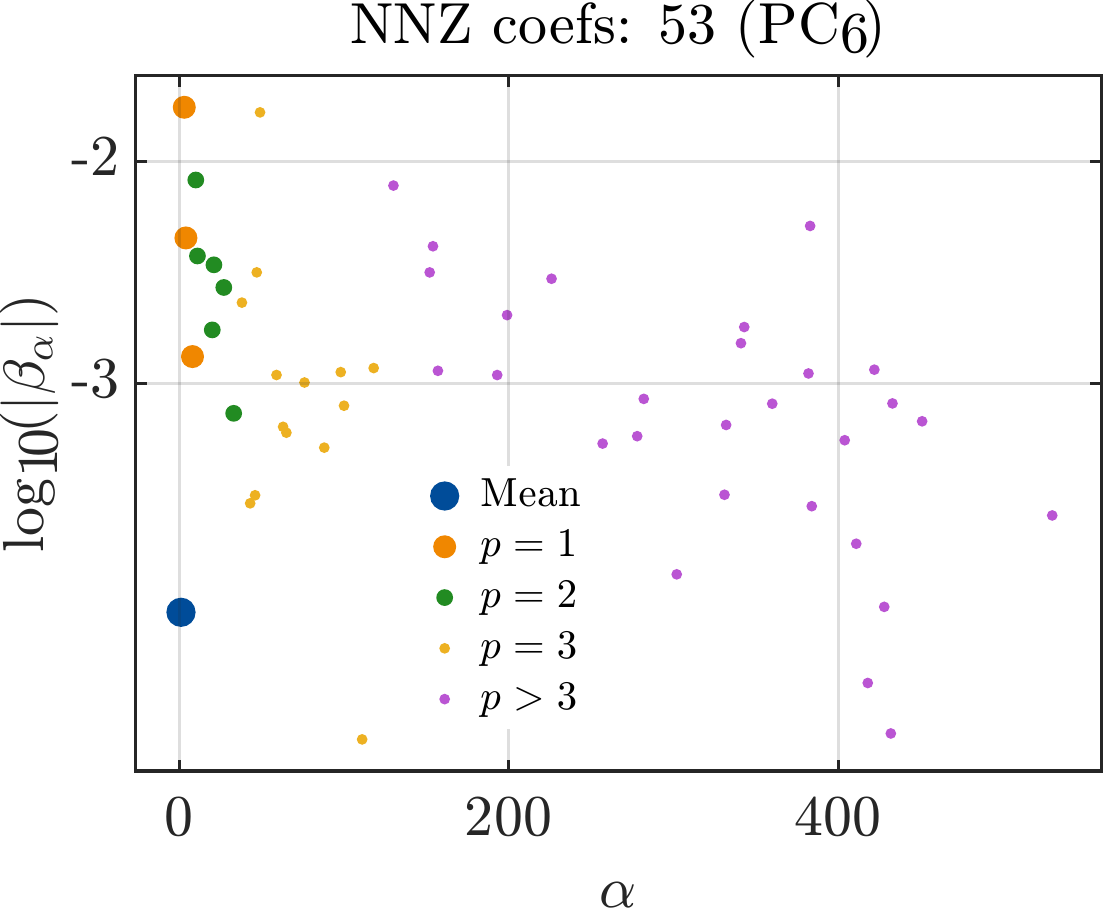}
  \vspace{5pt}
\end{subfigure}
\begin{subfigure}{0.33\textwidth}
  \centering
  \includegraphics[width=\linewidth]{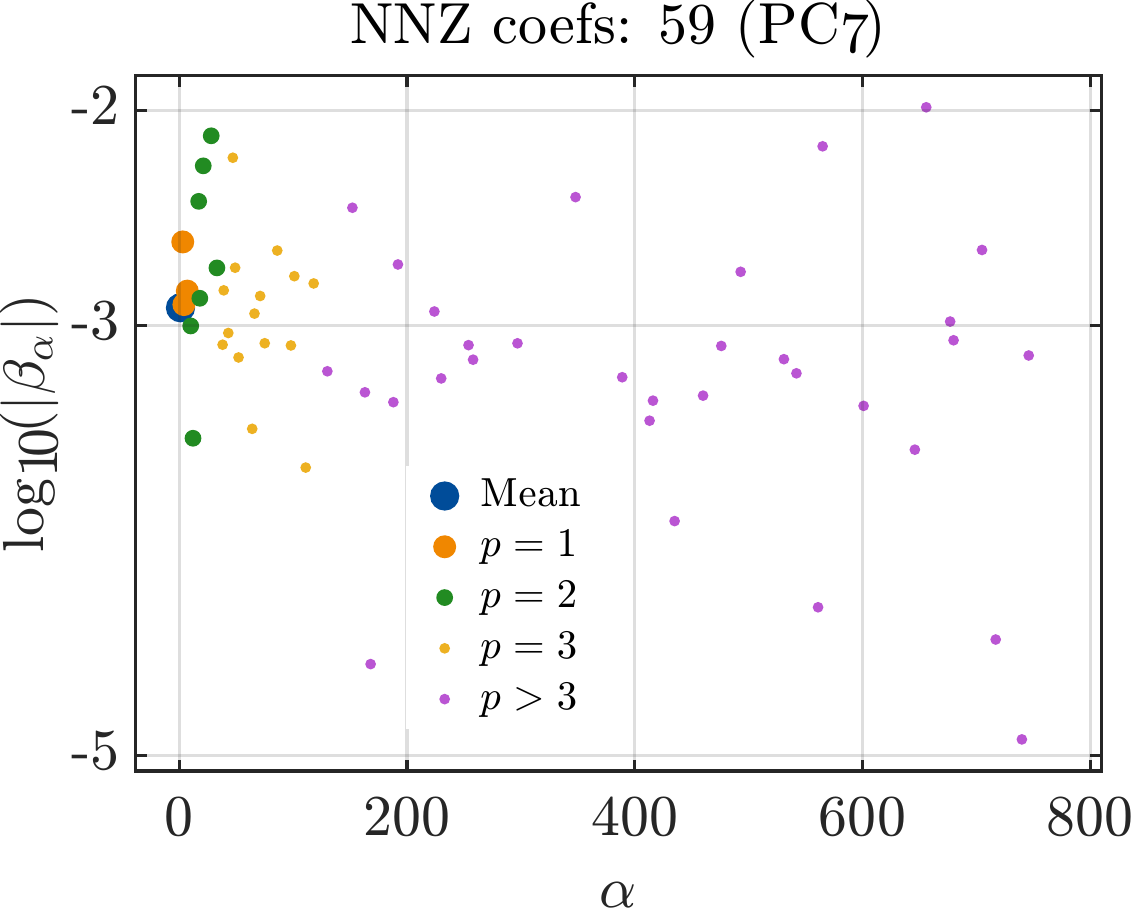}
\end{subfigure}%
\hfill
\begin{subfigure}{0.33\textwidth}
  \centering
  \includegraphics[width=\linewidth]{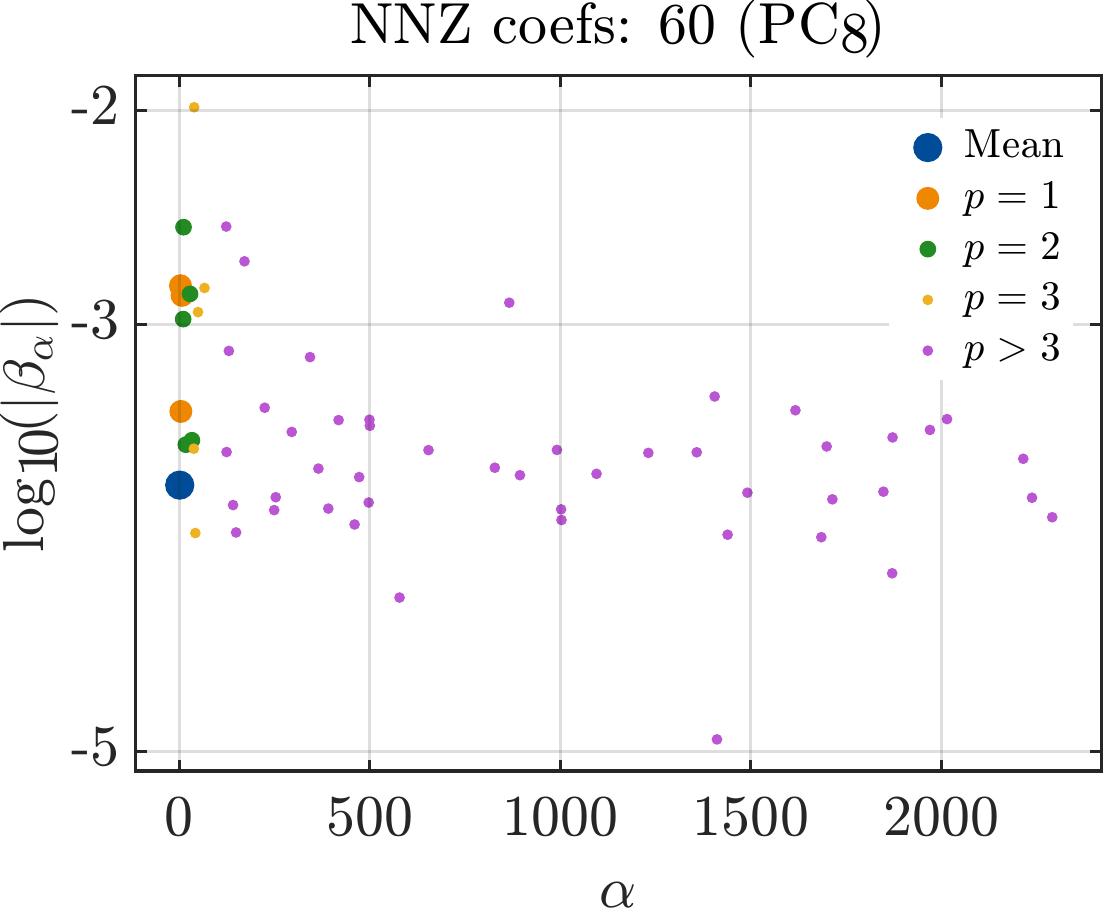}
\end{subfigure}%
\hfill
\begin{subfigure}{0.33\textwidth}
  \centering
  \includegraphics[width=\linewidth]{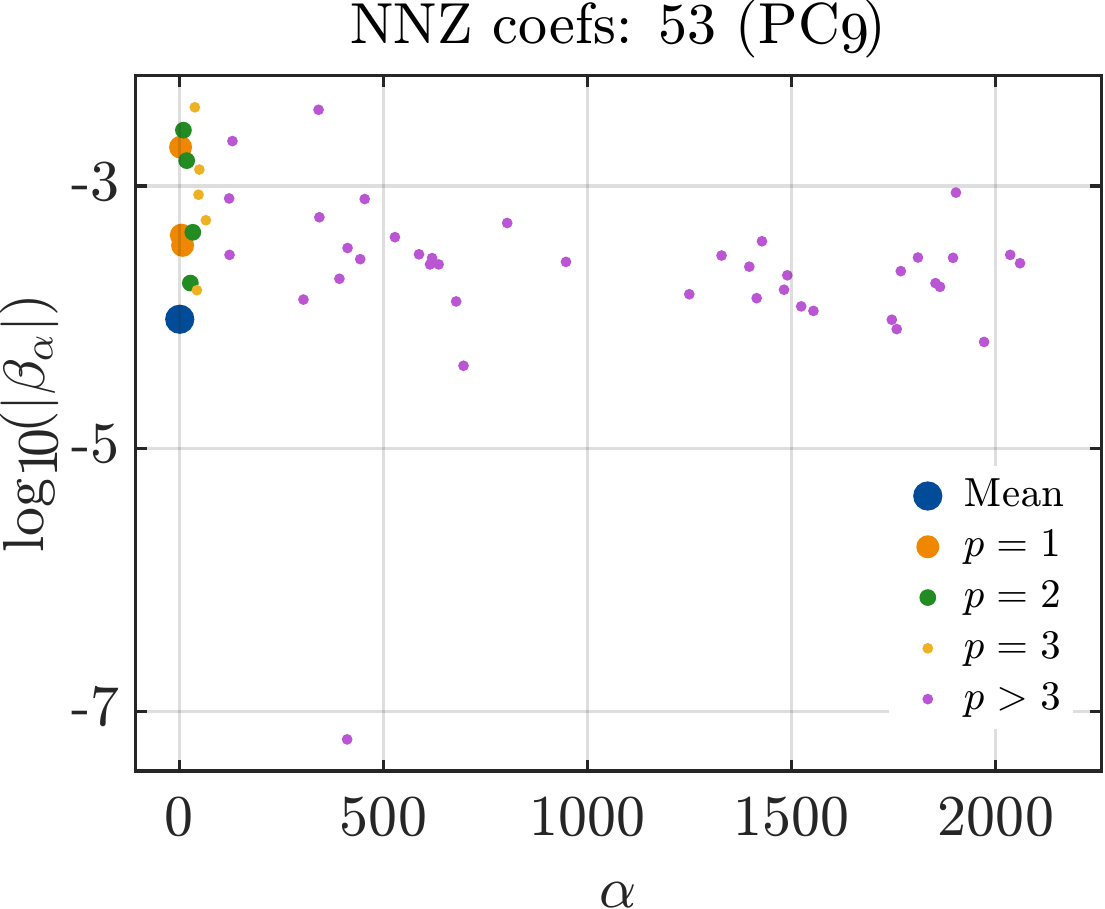}
\end{subfigure}

\caption{Figures of the log-scaled spectra of PCE coefficients obtained through sparse regression using the LARS method for each  principal component.}
\label{fig:NNZ}
\end{figure*}

In Fig.\ \ref{fig:PCBasis} we plot the mean of the dataset $\mathcal{D}$ alongside the nine principal components, which are derived by preserving $99.999\%$ of the variance in $\mathcal{D}$. These components are presented as functions of the wavenumbers at $z=0$.
Fig.\ \ref{fig:NNZ} illustrates the magnitude of the PCE coefficients, represented as $\log_{10}(|\beta_\alpha|)$, plotted against their enumeration index ($\alpha$) for all PCE expansions of the weights of principal components considered in constructing the \texttt{k-evolution}-based emulator.

\label{lastpage}

\printnoidxglossary[type=codes,style=tree, title=Glossary of Codes]
\printnoidxglossary[type=acronyms,style=tree, title=Glossary of Acronyms]

\end{document}